\documentclass{emulateapj}
\usepackage{graphicx}
\usepackage{amsmath}
\usepackage{lscape}
\usepackage{color}\def\red#1{}\def\blue#1{}

\newcommand{\feoh}{[{\rm Fe} / {\rm H}]}
\newcommand{\msun}{\, M_\odot}
\newcommand{\Zsun}{\, Z_\odot}
\newcommand{\mmd}{M_{\rm md}}
\newcommand{\dm}{\Delta_{\rm M}}
\newcommand{\Tv}{T_{\rm vir}}

\def\mhyph{\, \mathchar`- \, }

\def\abra#1#2{[{\rm #1}/ {\rm #2}]}
\def\Y#1{Y_{\rm #1}}

\def\rps{{\it r}-process }
\def\rdemp{{\it r}-deficient EMP }
\def\nor{no-{\it r} EMP }

\def\rsource{{\it r}-process source }
\def\BaEu{Ba AND Eu}

\shorttitle{\rps in the early Galaxy}
\shortauthors{Komiya et al.}

\begin{document}
\title{The New Model of Chemical Evolution of {\it r}-process Elements \\ Based on The Hierarchical Galaxy Formation I: \BaEu}

\author{Yutaka Komiya\altaffilmark{1}, Shimako Yamada\altaffilmark{2}, Takuma Suda\altaffilmark{1} and Masayuki Y. Fujimoto\altaffilmark{3,4}}
\altaffiltext{1}{National Astronomical Observatory of Japan, Osawa, Mitaka, Tokyo, Japan}
\altaffiltext{2}{Department of Cosmoscience, Hokkaido University, Sapporo, Hokkaido 060-0810, Japan}
\altaffiltext{3}{Nuclear reaction data center, Graduate School of Science, Hokkaido University, Sapporo, Hokkaido 060-0810}
\altaffiltext{4}{Visiting researcher, Faculty of Engineering, Hokkai-gakuen University, Sapporo, Hokkaido 062-8605}

\begin{abstract}

We investigate the chemical enrichment of \rps elements in the early evolutionary stages of the Milky Way halo within the framework of hierarchical galaxy formation using a semi-analytic merger tree.   
In this paper, we focus on heavy \rps elements, Ba and Eu, of extremely metal-poor (EMP) stars and give constraints on their astronomical sites. 
Our models take into account changes of the surface abundances of EMP stars by the accretion of interstellar matter (ISM). 
We also consider metal-enrichment of intergalactic medium (IGM) by galactic winds and the resultant pre-enrichment of proto-galaxies. 
The trend and scatter of the observed \rps abundances are well reproduced by our hierarchical model with $\sim 10\%$ of core-collapse supernovae in low-mass end ($\sim10\msun$) as a dominant \rps source and the star formation efficiency of $\sim 10^{-10} \hbox{yr}^{-1}$. 
For neutron star mergers as an \rps source, their coalescence timescale has to be $ \sim 10^7$yrs, and the event rates  $\sim 100$ times larger than currently observed in the Galaxy.    
We find that the accretion of ISM is a dominant source of \rps elements for stars with $\abra{Ba}{H} < -3.5$. 
In this model, a majority of stars at $\feoh < -3$ are formed without \rps elements but their surfaces are polluted by the ISM accretion. 
The pre-enrichment affects $\sim 4\%$ of proto-galaxies, and yet, is surpassed by the ISM accretion in the surface of EMP stars. 
\end{abstract}


\section{Introduction}\label{introS}

Extremely metal-poor (EMP) stars in the Galactic halo are the promising probes to reveal the early stages of the Galaxy formation since they are the very early generations of stars in terms of chemical evolution.  
In the cold dark matter (CDM) paradigm, galaxies are formed through mergers of proto-galaxies. 
The first generation of stars are thought to be born in mini-halos of total mass $M_{\rm h}\sim 10^6\msun$ \citep[e.g.][]{Tegmark97, Yoshida03}. 
EMP stars are also thought to be formed in small forming galaxies. 
We formulate a chemical evolution model in the framework of the hierarchical galaxy formation in order to investigate the metal enrichment history of the early Milky Way and origin of the elemental abundances of EMP stars. 

In this paper, we refer to stars with $\feoh\lesssim -2.5$ as EMP stars.  
As shown in our previous study \citep[][hereafter Paper~I]{Komiya10}, EMP stars have observational and theoretical peculiarities that distinguish them from more metal-rich Population~II stars (see Section 2.1 of Paper~I, for details). 
Especially, it is argued that the initial mass function (IMF) of EMP stars is different from more metal-rich stars and peaks around $10\msun$\citep{Komiya07, Komiya09}. 
A transition to the present-day IMF is thought to occur around $\feoh \simeq -2$ \citep{Suda11, Suda13, Yamada13}. 
We refer to the mother stellar population with $\feoh\lesssim -2.5$ including massive stars as EMP population and low mass survivors with nuclear burning now as EMP survivors, respectively. 
More metal-deficient objects with $\feoh<-4$ and $\feoh<-5$ are referred to as ultra metal-poor (UMP) stars and hyper metal-poor (HMP) stars, respectively. 
Only 7 UMP/HMP stars have been identified to date. 

It is known that abundances of heavy elements beyond the iron peak, such as Sr, Ba, and Eu, show large scatters with $\sim 2$ dex or more at $\feoh\lesssim-2.5$ \citep[e.g.][]{McWilliam95, Honda04}. 
Especially, stars with large enhancements of \rps elements with $\abra{Eu}{Fe}>+1$ are observed and referred to as $r$-II stars. 
These heavy elements are synthesized by the slow or rapid neutron capture processes; the {\it s}-process or {\it r}-process. 
The contribution of the {\it s}-process is thought to be negligible for EMP stars, however, since EMP stars have been formed before the first intermediate-mass stars, which are main {\it s}-process sites, end their lives.
In fact, the observed Ba/Eu ratios for EMP survivors are consistent with the pure \rps ratio \citep{McWilliam98, Burris00}. 
Accordingly, the \rps is the dominant source of these heavy elements for EMP survivors except for those born as the members of binaries with intermediate-massive stars as primary stars and enhanced with carbon and $s$-process elements through the mass transfer.  
In this paper, we focus on EMP stars without the carbon enhancement and regard Ba as \rps elements in addition to Eu.
For the heavier \rps elements ($Z\geq56$, such as Ba, Eu), the abundance pattern observed for these EMP survivors matches the scaled-solar \rps pattern \citep[e.g.][]{Francois07}. 
For the lighter neutron-capture elements ($Z<56$, such as Sr, Y, Zr), on the other hand, the enhancement relative to the heavier elements is observed for some EMP survivors \citep{Johnson02}. 
This may indicate that there are two (or more) \rps sources \citep{Travaglio04}. 
\citet{Aoki13} tried to reproduce the observed abundance scatter of $\abra{Sr}{Ba}$ using chemical evolution models with the \rps yields they proposed \citep{Boyd12}. 
In this paper, we investigate only Ba and Eu as representative of heavier \rps elements. 
The lighter \rps elements will be investigated in the forthcoming paper. 

The astronomical origin of the \rps elements is a long-standing mystery. 
Some probable sites have been suggested in association with supernova (SN) explosion.  
\citet{Mathews90} argued low-mass type~II SNe from the delayed production ofEu with relative to iron in EMP stars, and \citet{Wheeler98} argued the electron-capture SNe (ECSNe) of stars with O-Ne-Mg cores as a promising \rps site. 
A nucleosynthetic study by \citet{Wanajo03} showed that an ECSN yields a large amount of \rps elements when artificial large explosion energy is assumed. 
\red{ However, \citet{Wanajo11a} argue that ECSN cannot produce heavy \rps elements unless extremely neutron-rich material is assumed. }
Another possible site is neutrino-driven wind from proto neutron-stars (NSs) at the explosion of stars with $>20\msun$ \citep{Woosley92}. 
However, recent studies indicated the neutrino wind to be proton-rich for many seconds and heavier \rps elements are not synthesized \citep[see][for a review]{Thielemann11}.  

The coalescence of NS binary is also argued as a promising site of the \rps \citep[e.g.][]{Rosswog99}. 
The $r$-process nucleosynthesis in the tidally stripped material is shown to produce a nearly solar abundance pattern above mass number 130 \citep{Bauswein13}. 
Recent computations of nucleosynthesis in the neutrino-driven wind from the accretion disc around a black hole, formed as a remnant of binary NS merger succeed in representing a solar \rps abundance pattern \citep[e.g.][]{Wanajo12}. 

For all these theoretical models of \rps nucleosynthesis, there are still large uncertainties. 
$R$-process element yields are sensitive to an electron fraction at an \rps site and a thermal history of the ejected matter. 
Since the explosion mechanisms of core-collapse SNe (CCSNe) have not yet been revealed, theoretical models for nucleosynthesis at the explosion also suffer large uncertainty. 
Additionally, the \rps yields are also to depend on the three-dimensional mixing process and the neutrino transport during explosions \citep{Wanajo11a, Arcones11}, which are not well understood. 
Therefore, it is important to derive constraints for the astronomical origin of \rps elements from the viewpoint of chemical evolution. 

\red{The \rps source has been discussed from the perspective of chemical evolution. }
Using one-zone chemical evolution models, it have been attempted to estimate the mass range of the \rps sites in massive stars that can better reproduce the trend of observed data \citep[e.g.][]{Mathews92, Travaglio99, Cescutti06}. 
These studies suggest that both Eu and Ba in EMP survivors originate mainly from low mass CCSNe. 
Some authors investigate the abundance scatter of \rps elements taking into account that SN ejecta is not well mixed into the interstellar matter (ISM). 
\citet{Ishimaru99} calculated the evolution of ISM in the Galactic halo with a one-zone model assuming that a formed star has the mass averaged chemical composition of the ejecta of a SN that triggers the star formation and the ``snowplowed'' ISM, swept up by its remnant. 
\citet{Tsujimoto99} investigated the enrichment of Eu in the Galactic halo assuming a SN-triggered star formation scenario. 
\citet{Travaglio01} represented the halo gas by an ensemble of clouds of different chemical composition uniform in each of them but different from each other, and assume that the clouds coalesce randomly and fragment by starbursts. 
\citet{Fields02} gave a simple analytic expression for the abundance scatter versus $\feoh$. 
\citet{Argast04} presented an inhomogeneous chemical enrichment study and compared the scenario that considers neutron star merger as a major \rps source to that in which CCSNe act as a dominant \rps site. 
They ruled out the NS merger scenario since it leads to too low $\abra{\it r}{Fe}$ at $\feoh<-3$ and to too large scatter at all the metallicity range. 
\citet{Cescutti08} developed a stochastic chemical evolution model in which the halo consists of 100 independent regions. 
They argue that the production for \rps elements extends to the range between 12 and 30$\msun$. 

All these previous studies predict a considerable scatter of neutron capture elements.  
And yet, they did not consider the galaxy formation process in context of the concordance cosmology. 
We consider the merging history of mini-halos based on the $\Lambda$CDM cosmology, and a proto-galaxy in each mini-halo evolves independently along the merger tree in our model. 
In Paper~I, we build a merger tree based on the extended Press-Schechter theory, and demonstrated that our hierarchical model can reproduce the metallicity distribution of the halo stars. 
The enrichment of $\alpha$-elements and iron group elements were investigated in Komiya (2011, hereafter Paper~II). 
In this paper, we compute the enrichment history of \rps elements by updating our hierarchical chemical evolution model. 

\red{  Some previous studies have considered the metallicity distribution function (MDF) of the Milky Way halo in the hierarchical merging paradigm. 
\citet{Prantzos08} accounted for the Milky Way halo MDF as sum of dwarf galaxies without constructing a merger tree. 
\citet{Calura09} studied chemical evolution using semi-analytic galaxy formation models but their models cannot be applicable for EMP stars. 
\citet{Tumlinson06} and \citet{Salvadori06} discussed the MDF of the Milky Way halo using hierarchical merger trees. 
All these studies did not consider \rps elements. }

One of novelties of this study is that we consider the change of surface abundance of EMP survivors by the accretion of ISM in the hierarchical model.  
We follow the accretion of ISM and changes in the surface abundances along with the chemical and dynamical evolution of the Galaxy. 
We present the predicted distributions both of the intrinsic abundances, $\abra{X}{Fe}_{\rm i}$, and the surface abundances, $\abra{X}{Fe}_{\rm s}$, of EMP survivors. 
To evaluate the effect of ISM accretion, we have to take into account not only chemical evolution but also growth history of halo mass since the ISM accretion rate is strongly dependent on the velocity of stars. 

In addition, unlike the previous studies, we explicitly take into account EMP stars without \rps elements for completeness. 
It is thought that there are SNe which eject iron but not \rps elements. 
Out of the gas polluted by the ejecta from these SNe, therefore, EMP stars without \rps elements are formed. 
We referred to EMP stars which contain no \rps elements in their interior ($\abra{r}{H}_{\rm i}=-\infty$) as \nor stars and consider them in this paper, although they are neglected in most of the previous inhomogeneous chemical evolution studies. 
Their surfaces are polluted by the accretion of ISM with \rps elements but some EMP survivors remain with very low \rps abundance in the surface. 
We refer to EMP stars of the surface abundances of \rps elements below the detection limit of current observations, i.e., with $\abra{Ba}{H}_{\rm s}<-5.5$, as \rdemp stars in this paper. 
We estimate the number of \nor stars and \rdemp stars, and the distribution of their surface abundance, $\abra{r}{Fe}_{\rm s}$. 

Another improvement in this paper is our examination of gas outflow from proto-galaxies by SNe and pre-enrichment of intergalactic medium (IGM) by the outflow.  
Outflow from mini-halos is effective due to their small gravitational potential. 
We follow inhomogeneous metal enrichment process of the Galactic IGM along a merger tree. 
Metal enrichment of IGM by galactic winds has been investigated \citep{Aguirre01,Bertone07}, but outflow from mini-halos which is triggered by individual Pop.III stars in the early universe is not explicitly considered in these previous studies. 
Furthermore, the effect of pre-enrichment on the elemental abundances of EMP stars is yet to be investigated. 
We focus on the IGM pre-enrichment by early generations of stars including Pop~III stars, and locate their signature on EMP stars.  

The purpose of this work is twofold. 
First, we explore the chemical evolution process of \rps elements using the new model, based on the concordance $\Lambda$CDM cosmology, and show the diversity of the chemical evolution of proto-galaxies. 
Especially, we evaluate the effects of the ISM accretion on EMP survivors and the IGM pre-enrichment by the outflow from proto-galaxies. 
Second, we draw constraints on the \rps site(s) from the comparison of predicted abundance distributions with observational properties such as the metallicity dependences of mean enrichment and scatter of their abundances, and the frequencies of $r$-II stars and \rdemp stars. 

In Section \ref{modelS}, we describe our hierarchical chemical evolution model and assumptions. 
In Section \ref{resultS}, we show the computation results. 
The detailed evolution process for our fiducial model is described in section \ref{scenarioS} and parameter dependences are shown in Section \ref{paramS}. 
We present the comparison with observations and give constraints on the \rps site in Section \ref{obsS}. 
In Section~\ref{NSMS}, we discuss the other scenario with the NS mergers as the \rps site.  
Summary and conclusions are presented in Section~\ref{concludeS}.

\section{Computation Method and Assumptions}\label{modelS}

We study the chemical enrichment of \rps elements for EMP stars in the Milky Way halo by using a semi-analytic merger tree constructed within the framework of the $\Lambda$CDM cosmology. 
   We refer to the baryonic component in a (mini-)halo as a proto-galaxy. 

\begin{figure*}
\includegraphics[width=\textwidth]{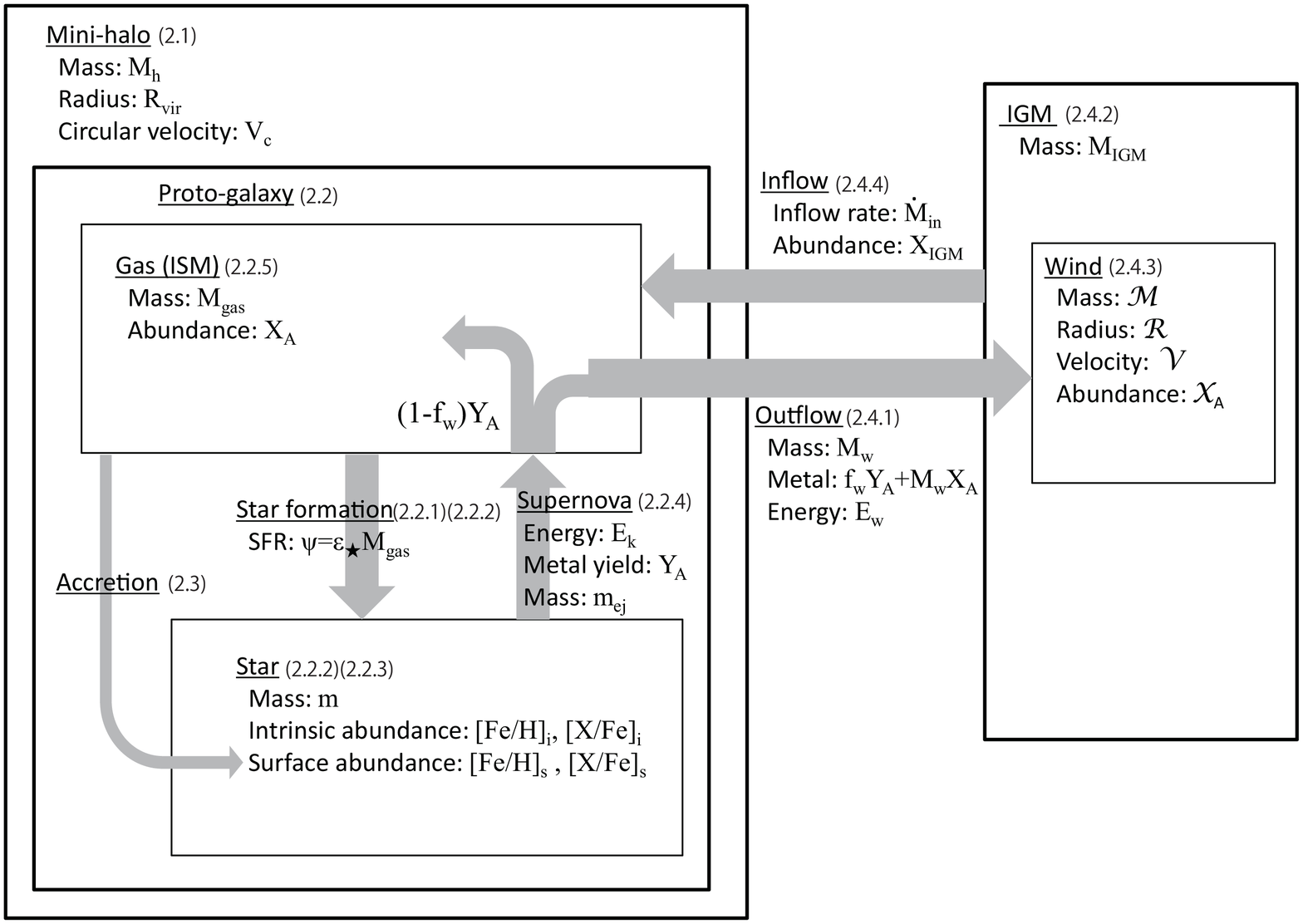}
\caption{The schematic diagram of the chemical evolution model for the proto-galaxies and the inter-galactic medium (IGM) surrounding them.  
   The formation and merging history of proto-galaxies and the infall rate ($\dot{M}_{\rm in}$) of IGM onto them are described by the merger tree, build by the extended Press-Schechter method. 
   Stars are formed ($\psi$) in the proto-galaxies and mass ($m$) of individual stars are set randomly following the IMF. 
   Massive stars become SNe and eject metals ($Y_{A}$) to the interstellar matter (ISM).  
   The kinetic energy ($E_{\rm k}$) of SN explosion (and HII region) give rise to galactic outflow, and eject mass ($M_{\rm w}$) and metals from proto-galaxies. 
   In our model, the matter ejected from the proto-galaxies forms galactic winds, and they expand into and mixed with the IGM. 
   The surface abundances of stars are changed after their formation due to the accretion of ISM. 
   Numbers in the figure show sub-sections that describe the details. 
}\label{diagram}
\end{figure*}

Figure~\ref{diagram} is the schematic diagram for the flow of gas and metal for the proto-galaxies and the IGM.  
  Stars are formed in the proto-galaxies and the masses of individual stars are set randomly according to the IMF. 
   Massive stars explode as SNe and eject metals to the ISM.  
   The kinetic energy of SN explosion (and HII region) gives rise to galactic outflow, and ejects mass and metals from proto-galaxies. 
   The matter ejected from the proto-galaxies forms galactic winds, and it expands into and mixed with the surrounding IGM.  
 \red{  Mini-h}alos which collapse later may contain IGM polluted with metals ejected by the galactic wind and have different initial abundances \red{from the pristine gas}.  
   The surface abundances of stars are changed after their birth owing to the accretion of the ISM polluted with metals. 
   We describe the details of our models in the following 

\subsection{Merger Trees}

The basic assumptions of our hierarchical chemical evolution model are the same as used in Paper I and Paper II. 
We build merger trees using the ``$N$-branch trees with accretion'' method of \citet{SK99} based on the extended Press-Schechter theory \citep{LC93}. 
We use the standard $\Lambda$CDM cosmology with $\Omega_\Lambda=0.7, \Omega_{\rm M}=0.3$, and $\Omega_{\rm b}=0.045$.
The trees give the mass accretion rate to each mini-halo $\dot M_{\rm h}$ and the merging history of mini-halos. 
The total mass of the Milky Way halo is taken to be $M_{\rm MW} = 10^{12}\msun$ at $z=0$. 
The mass resolution is determined to give the minimum virial temperature at $\Tv{}_{,m}=10^3$K.

\subsection{The Chemical Evolution of Proto-galaxies}\label{model1S}

The chemical abundance of $\sim10^5$ proto-galaxies in the merger trees are traced in the computation. 
We set the initial mass of proto-galaxy to be $M_{\rm gas} = M_{\rm h} \Omega_{\rm b}/\Omega_{\rm M}$ and the gas inflow rate to be $\dot M_{\rm in}= \dot M_{\rm h} \Omega_{\rm b}/\Omega_{\rm M}$.
For the redshift $z<10$, mini-halos with $\Tv<10^4$K cannot accumulate gas, $\dot M_{\rm in}=0$, due to reionization of the universe \citep[e.g.][]{Komatsu09}.

\subsubsection{The Star Formation Rate}

The star formation starts in the mini-halos with $\Tv\ge10^3$K at the beginning, but is suppressed in mini-halos formed at $z<20$ and $\Tv<10^4$K due to the photo-dissociation of H$_2$ molecules by Lyman-Werner background radiation \citep[e.g.][]{Ricotti02}. 
The star formation rate, $\psi$, is assumed to be proportional to gas mass; $\psi=\epsilon_{\star}M_{\rm gas}$, where $\epsilon_{\star}$ is star formation efficiency (SFE). 
We may discuss the dependence on the SFE in Section~\ref{resultSFES}. 
In our fiducial model, the SFE turns out to be $\epsilon_{\star}=10^{-10}$/yr, similar to the present day SFE of the Milky Way \citep[e.g., the review by][]{Kennicutt12}. 
Stars are formed discretely and independently in each proto-galaxy. 
All the individual metal-poor stars with $Z\le 0.1\Zsun$ are registered in our computation.  
For metal-rich ($Z>0.1\Zsun$) proto-galaxies, we register aggregations of low-mass stars and high-mass stars for each timestep of $3 \times 10^6$ yr, instead of individual stars, and the chemical evolution is computed with the use of the IMF-averaged yields.

\subsubsection{The Initial Mass Function (IMF)}

The mass of each metal-poor star is set randomly according to the IMF. 
We adopt a lognormal IMF; 
\begin{equation}
\xi (\log m)\propto 
\exp \left[ -\frac{ \{ \log (m/ \mmd) \} ^2}{2\times \dm^2} \right]   
\label{eq:IMF-lognormal}
\end{equation}  
which is connected smoothly to a power law distribution with the same power of the Salpeter IMF for high-mass end, i.e., $\xi \propto m^{-1.35}$. 
For stars with the metal abundance of $\abra{Z}{H}>-6$, we set $\mmd = 10\msun$ and $\dm = 0.4$ according to our previous results, which is derived from the statistics of carbon-enhanced EMP stars \citep{Komiya07, Komiya09}. 
  In addition, we assume that a half of stars are born in binary with a lower-mass companion specified by a flat mass-ratio distribution function. 
 This high-mass IMF with binary contribution is shown to reproduce the observed total number of EMP survivors by Hamburg/ESO survey, in addition to the large fraction ($10-25\%$) of carbon-enhanced stars among EMP survivors \citep[][see also Fig.~\ref{MDF} below]{Komiya09}. 
A low mass IMF such as the standard Salpeter IMF predicts $\sim100$ times as large total number of EMP survivors as observed in addition to a small fraction ($\sim1\%$) of carbon-enhanced stars. 
We also present the results using a low-mass IMF by \citet{Chabrier03} for the Galactic spheroid in Section \ref{resultIMFS}. 
   As long as concerned with \rps elements, the predicted abundance distributions result similar for the both IMFs as shown below.  
   Note, however, that the low-mass IMF demands the star formation efficiency larger by $\sim 5$ times to birth the same number of massive stars and promote the chemical enrichment in the same time scale. 
We do not consider the transition of the IMF in this study \blue{(but see \citet{Suda13} and \citet{Yamada13}).}

\subsubsection{Population~III star}

In this paper, we regard stars with the metal abundance below $\abra{Z}{H} = -6$ as Population~III (Pop.~III) stars since it is argued that the dust cooling enables low-mass fragmentation and lowers the typical mass of stars formed above $\abra{Z}{H} = -6$ \citep{Omukai05, Schneider06}.
We adopt higher $\mmd$, i.e., $\mmd=200\msun$ for the first generation Pop.~III.1 stars and $\mmd=40\msun$ for the Pop.~III.2 stars formed of the ionized gas \citep[e.g.,][]{Yoshida07}.  
We compute the switchover from Pop.~III.1 to Pop.~III.2 stars by a statistical treatment of the ionization of IGM by massive stars (see Paper I for details).  
We adopt the same binary frequency and mass ratio distribution as EMP stars. 
A few low-mass Pop.~III survivors are formed mostly as the secondary members of binaries. 
The first massive Pop.~III star, if formed, prohibits the star formation in its host proto-galaxy until it explodes as a SN. 

\red{
The assumptions adopted can reproduce the low-metallicity tail of the MDF for the halo stars, as shown in Figure~\ref{MDF}. 
If we adopt the same IMF with EMP stars, too much Pop.~III survivors are predicted as shown in Paper~I. }
\red{For the \rps element abundance distribution, however, the Pop.~III IMF have little impact as seen from our results. }

\subsubsection{The Stellar Yields}

The iron yield, $Y_{\rm Fe}$, and the explosion energy, $E_{\rm SN}$, of iron core collapsed SN (FeCCSNe) are adopted from the mass and metallicity dependent theoretical results of \citet{Kobayashi06}. 
Stars with initial mass at 10--$40\msun$ explode as FeCCSNe. 
A half of the stars of initial mass $m > 20 \msun$ are taken to explode as hypernovae with $E_{\rm SN}>10^{52}$ erg. 
For the yield of pair-instability SNe, the results of \citet{Umeda02} are used. 
For the ECSNe, we adopt the result of \citet{Wanajo09}. 
Stellar lifetimes are specified following \citet{Schaerer02}. 
\red{For type~Ia SN, we assume that $9\%$ of binaries with primary mass at $2\msun <m< 9\msun$ become type~Ia SNe at the end of lifetime of their secondary companions \citep{Greggio05}. 
We adopt W7 model in \citet{Nomoto84} for the metal yield of type~Ia SNe. 
}

As for the \rps elements, there is no reliable theoretical model both for the sites and yields.  
   In this paper, therefore, we compute the chemical evolution of \rps elements with the relevant assumptions (see Section~\ref{MrS}) in an attempt to draw the constraints on them through the comparisons of our computation results with the observations (see Section~\ref{obsS}).

\subsubsection{The Variations of Abundances of Gas}

The gas of each proto-galaxy is assumed to be chemically homogeneous, which holds good at least for mini-halos of baryon mass of $10^5 - 10^6 \msun$ \citep[e.g.,][]{Machida05}.  
The mass, $M_{{\rm gas},n}$, of gas and the abundances, $X_{A,n}$, of chemical element $A$ in an $n$-th proto-galaxy evolve as follows,
\begin{equation}
\frac{d}{dt} M_{{\rm gas},n} = \dot M_{{\rm in},n} + \sum_i \{m_{\rm ej} (m_{i,n})- M_{{\rm w},i,n} \}\delta(t-t_i)  - \epsilon_\star M_{{\rm gas},n} 
\end{equation}
\begin{align}
\frac{d}{dt} (M_{{\rm gas},n} & X_{A,n}) = \dot M_{{\rm in},n} X_{{\rm IGM},A,n}  \notag\\
& + \sum_i \lbrace (1-f_{{\rm w},i,n})  Y_{A}(m_{i,n})- M_{{\rm w},i,n} X_{A,n}\rbrace \delta(t-t_i) \notag\\
& - \epsilon_\star M_{{\rm gas},n} X_{A,n}.   
\end{align}
Here the first and the last terms in the right-hand sides are due to gas infall and to the star formation, respectively, where $X_{{\rm IGM},A, n}$ is the element abundance of IGM around the $n$-th halo (see Sec.~\ref{XigmS} for details): 
The second terms describe contributions from the $i$-th SNe that occurred at $t_i$ in the $n$-th proto-galaxy, 
 where $m_{i, n}$ is the mass of the progenitor star, $m_{\rm ej}$ and $Y_{A}$ are the ejected mass and the element yield, respectively, of SNe as functions of $m_{i, n}$:   
 $M_{{\rm w},i,n}$ and $f_{{\rm w},i,n}$ are, respectively, the mass of the gas blown away from the $n$-th proto-galaxy and the fraction of SN yield lost by the galactic wind, triggered by the $i$-th SNe, which are determined as functions of $E_{\rm SN}, M_{\rm h}$ and $M_{\rm gas}$, as described in Section \ref{IGMmodelS} with the chemical enrichment of the Galactic IGM. 

\begin{table*}
\begin{center}
\caption{Parameters and their fiducial values}
\label{Tparam}

\begin{tabular}{|l|c|c|}
\hline
description	& parameter & fiducial value \\
\hline
\multicolumn{3}{|c|}{merger tree} \\
\hline
Total mass of the Galaxy (dark matter + baryon) & $M_{\rm MW}$	& $10^{12}\msun$ \\
Minimum virial temperature of the mini-halo & $\Tv{}_{,m}$	& $10^{3}$K \\
\hline
\multicolumn{3}{|c|}{star formation} \\
\hline
Star formation efficiency	&	$\epsilon_{\star}$	& $10^{-10} yr^{-1}$ \\
Median mass of Pop.~II, and Pop.~I stars &	$\mmd[\rm EMP]$	& $10\msun$ \\
Median mass of Pop.~III.1 stars &	$\mmd[\rm III.1]$	& $200\msun$ \\
Median mass of Pop.~III.2 stars &	$\mmd[\rm III.2]$	& $40\msun$ \\
Boundary metal abundance between Pop.~III and Pop.~II stars &	$Z_{\rm cr}$	& $10^{-6}\Zsun$ \\
\hline
\multicolumn{3}{|c|}{outflow} \\
\hline
SN explosion energy (normal SN)	&	$E_{\rm SN}$	&	$10^{51}$ erg  	\\
  \hspace{9em} (hypernova)	&		&	$(1-3) \times 10^{52}$ erg  \tablenotemark{a} 	\\
	\hspace{9em} (ECSN)	&		&	$ 10^{50}$ erg  	\\
Kinetic energy fraction among SN explosion energy	&	$\eta$ & 0.1 \\
Minimum fraction of kinetic energy to go to wind	&	$\epsilon_k$	& 0.1 \\
\hline
\end{tabular}
\tablenotetext{a}{Dependent on initial mass of stellar progenitor. See Kobayashi et al. for details.}
\end{center}
\end{table*}

\subsection{The Surface Pollution of EMP survivors by ISM Accretion}\label{polmodelS}

One of the important ingredients in our model is the surface pollution of EMP survivors by the accretion of inter stellar matter (ISM), enriched with metals by SN ejecta. 
As shown in our previous studies \citep[][Paper~I]{Yoshii81, Iben83, Suda04}, the ISM accretion can be the dominant source of the surface metal abundances for UMP/HMP stars. 
\red{In addition, the ISM accretion has large impact on the \rps element abundance distribution of EMP stars as shown later. }
We follow the changes of surface abundances for all the individual EMP survivors and stars with $\abra{Ba}{H}_{\rm i}\leqq-2.5$ in the computations. 

We assume the Bondi-Hoyle accretion. 

The accreted matter is well mixed in the surface convection zone of EMP stars. 
Mass of the surface convection zone is $0.2\msun$ for giant stars and $0.0035\msun$ for dwarf stars \citep{Fujimoto95}. 
\red{In the case of binaries, we simply assume that a half of accreted ISM settles onto each member. }

These assumptions are the same with Case~D in Paper~I. 
\red{In Paper~I, we showed that the ISM accretion can raise the surface abundance of low-mass Pop.~III giants through $\feoh \sim -5$. }

\red{
The surface pollution is most effective in the host proto-galaxies in which EMP stars are formed because of the small velocity of stars relative to ISM, as shown in Paper~I \citep[see also][]{Suda04}. 
The accretion rate reduces after the host proto-galaxies undergo the merger with larger proto-galaxies owing to the increase in the relative velocity. 
\citet{Frebel09} assessed the effect of the ISM accretion during disk crossing of halo stars and concluded that the accretion is negligible even for the HMP stars simply because they did not consider the change of stellar dynamics by the hierarchical galaxy formation process. 
In the current Milky Way, the accretion rate is much lower due to the large relative velocity of halo stars to the gas clouds. }

\red{
}

\subsection{Galactic Wind and The Metal Pre-Enrichment of IGM}\label{IGMmodelS}

The initial metallicity of proto-galaxies is not always zero owing to the metals blown off from older proto-galaxies. 
The metal pre-enrichment by the outflow from proto-galaxies can be important for the very early generations of stars. 
As well as proto-galaxies which form Milky-Way at $z=0$, we follow the chemical evolution of the intergalactic medium(IGM), in which the proto-galaxies are embedded.  
Out of the gas in the computation region with $M_{\rm MW}$, the baryonic component which resides outside of mini-halos are referred to as IGM in this paper. 
They are the sum of the gas which will accrete to the Milky Way and the matter which is blown-out from proto-galaxies. 

While we assume the chemically homogeneous IGM in Papers~I and II, we build an inhomogeneous model in this study. 
Some previous studies investigate the metal enrichment of intergalactic matter using the cosmological simulations of dark matter halo and the semi-analytical models of the galactic wind \citep[e.g.,][]{Aguirre01,Bertone07}, but they \red{are mostly concerned with the winds from galaxies of larger masses ($M \gtrsim 10^8\msun$), and consider neither the detailed element abundances nor effects of pre-enrichment on proto-galaxies. }
We focus on the metal pre-enrichment of IGM by the winds triggered by the first or early generations of stars in proto-galaxies of small masses and their nucleosynthetic signature on the abundances of EMP survivors.

\subsubsection{Outflow from Proto-Galaxies}

We apply the following formulation to the outflow of gas from the proto-galaxies.  
\red{ We consider the outflow triggered by an individual SN because the first SN is thought to be able to blow out the gas in their host mini-halo \citep{Machida05, Kitayama05, Greif07}.  }
The outflow is driven by the kinetic energy of SN explosion and of expansion of HII region. 
A portion, $\eta$, of SN explosion energy, $E_{\rm SN}$, is converted into the kinetic energy of gas shell, where we assume $\eta = 0.1$ \citep[e.g.][]{Machida05}. 
In addition to SN, HII regions, which are formed by massive stars, may also blow the gas away from proto-galaxies. 
We specify the kinetic energy of HII region, excited by a star with mass $m$, by $10^{48} \times (m/\msun) {\rm erg}$\citep[e.g.,][]{Kitayama04}. 
Thus, the total kinetic energy (SN + HII region) amounts to $E_{\rm k}=\eta E_{\rm SN}+10^{48} \times (m/\msun) {\rm erg}$. 

The outflow rate is thought to depend on the binding energy, $E_{\rm bin}$, of gas in the host proto-galaxy. 
\red{We define $E_{\rm bin}$ as the energy to expel the gas of mass $M_{\rm gas}$ from darkmatter halo of mass $M_{\rm h}$. }
 In particular, when $E_{\rm k}$ sufficiently exceeds $E_{\rm bin}$, all of the gas in the proto-galaxy will be blown-off from the host halo with almost all the kinetic energy, i.e., the energy, $E_{\rm w}$ and the mass load, $M_{\rm w}$, of the outflow are $E_{\rm w}=E_{\rm k}$ and $M_{\rm w}=M_{\rm gas}$, respectively. 
  On the other hand, when $E_{\rm k} \ll  E_{\rm bin}$, we may assume that a part, $\epsilon$, of the kinetic energy turns to the wind energy, i.e., $E_{\rm w} = \epsilon E_{\rm k}$. 
 
In our computation, we may adopt the following interpolation formulae for the energy and the mass load for the wind;
\begin{eqnarray}
E_{\rm w} &=& E_{\rm k} \frac{\epsilon +E_{\rm k}/E_{\rm bin}}{1+E_{\rm k}/E_{\rm bin}} \label{energyout} \\
M_{\rm w} &=& M_{\rm gas} \frac{E_{\rm w}}{E_{\rm bin}+E_{\rm w}} \label{massout}, 
\end{eqnarray}
   respectively. 
   Fraction, $f_{\rm w}$, of metal which is carried out by the wind among the metal yield of the SN is thought to be $M_{\rm w}/M_{\rm sw}$ when the metal is well mixed in the swept-up gas with mass $M_{\rm sw}$.  
   When $E_{\rm k} \gg E_{\rm bin}$, then, $f_{\rm w}$ is thought to be unity. 
   We adopt the following formula; 
\begin{equation}
f_{\rm w} =  \min \left( 1, \frac{M_{\rm w}/M_{\rm sw}+E_{\rm k}/E_{\rm bin}}{1+E_{\rm k}/E_{\rm bin}} \right),  \label{metalout}
\end{equation}
and $M_{\rm sw}$ is taken to be $M_{\rm sw}=10^5 \times (E_{\rm SN}/10^{51}$erg$)^{1/2} \msun $ \citep{Machida05}.

The current knowledge of mass and metal outflow rate from galaxies are still poor. 
   In the present study, we adopt $\epsilon = 0.1$ as a rule of thumb \citep[e.g.,][]{Fujita04}.  
   Our fiducial model gives average metallicity of $\feoh\sim-1$ for the Galactic IGM at z=0 \citep{Dunne03}. 
   We note that our results for abundance of EMP survivors may not be sensitive to the exact choice of the outflow parameters 
   since outflow from small mini-halos of $E_{\rm k} \gtrsim  E_{\rm bin}$ dominates the pre-enrichment at high redshift.

\subsubsection{IGM}
In this computation, IGM is the sum of the gas which will accrete to the Milky Way and the matter which is blown-out from proto-galaxies. 
The initial value of the total mass, $M_{\rm IGM}$, of the Galactic IGM is $M_{\rm MW}(\Omega_{\rm b}/\Omega_{\rm M})$ and it changes as follows; 
\begin{equation}
\frac{d}{dt} M_{\rm IGM} = \sum_n \left( \sum_i M_{{\rm w},i,n}\delta(t-t_i) - \dot M_{{\rm in},n} \right),  
\label{IGMmass}
\end{equation}
where $M_{{\rm w},i,n}$ is the mass, brown out by the $i$-th SN from the $n$-th halo, which is set by eq.~(\ref{massout}),  
and $\dot M_{{\rm in},n}$ is the infall rate to the $n$-th halos. 

\red{The density, $\rho_{\rm IGM}$, of the Galactic IGM is given by $M_{\rm IGM}/M_{\rm MW}$ times the averaged matter density of the universe. }

\subsubsection{Evolution of the Galactic Wind in the IGM}
\red{The outflow from proto-galaxies forms galactic winds and enriches the IGM with metal. 
The IGM enrichment can be a dominant source of metal for stars in the proto-galaxies formed with the gas pre-enriched by the galactic wind. 
We follow the patchy inhomogeneous metal enrichment process of the IGM by winds from proto-galaxies. 
}

The evolution of the wind from proto-galaxies and the metal enrichment of the Galactic IGM is formulated as follows. 
The first massive star in each mini-halo forms a galactic wind. 
Gas and metal mass of the wind are described in eqs.~(\ref{massout}) and (\ref{metalout}), respectively.
The subsequent SNe in the mini-halo \red{ also blow-out gas and metal in the same way. 
The mass and metal ejected by the second and later SNe are added to the existent galactic wind formed by the first SN and are assumed to be well mixed with those previously ejected.  }
We also assume an additional momentum, $\sqrt{2E_{\rm w}M_{\rm w}}$, to the wind. 

We may describe the evolution of the galactic wind in the IGM by assuming the snowplow shell-model with the momentum conservation in a spherical symmetry.  
The growth of the mass, ${\cal M}$, the momentum ${\cal M V}$, and the chemical abundance, ${\cal X}$, of the wind ejected from $n$-th halo are written in the form;
\begin{align}
\label{windmass}
\frac{d}{dt} {\cal M}_{n} &= ({\cal V}_n-{\cal R}_n H_r) 4\pi {\cal R}_n^2 \rho_{\rm IGM}  \notag \\
& + \sum_i M_{{\rm w},i,n}\delta(t-t_i) - \sum_{\{m|m \in S_n\}} \dot M_{{\rm in},m} ,
\end{align}
\begin{align}
\label{windmoment}
\frac{d}{dt} &( {\cal M}_{n} {\cal V}_n) = ({\cal V}_n-{\cal R}_n H_r) 4\pi {\cal R}_n^2 \rho_{\rm IGM} \cdot {\cal R}_n H_r \notag \\
& + \sum_i \sqrt{2 E_{{\rm w},i,n}M_{{\rm w},i,n}} \delta(t-t_i) - \sum_{\{m|m \in S_n\}} \dot M_{{\rm in},m} {\cal V}_n .
\end{align}
\begin{align}
\label{windabund}
\frac{d}{dt} ({\cal M}_{n} {\cal X}_{A,n}) &= \sum_i (M_{{\rm w},i,n}X_{A,n} + f_{{\rm w},i,n} Y_{A,i,n}) \delta(t-t_i) \notag \\
&- \sum_{\{m|m \in S_n\}}  \dot M_{{\rm in},m} {{\cal X}_{A,n} }. 
\end{align}
Here the physical quantities of winds are denoted by calligraphic font, and ${\cal R}$ and ${\cal V}(\equiv \dot {\cal R})$ are the outer radius and expansion velocity of the wind shell, respectively: and $H_r$ is the Hubble parameter. 
The initial radius is set at the virial radius of the progenitor mini-halo. 
The first terms in the right sides of eqs.(\ref{windmass}) and (\ref{windmoment}) describe the loading of the swept-up IGM. 
The second terms of eqs.(\ref{windmass}) and (\ref{windmoment}) and the first term of eq.(\ref{windabund}) are the gas mass, momentum, and metal mass of outflow by the $i$-th SNe in the progenitor halo, 
\red{where $E_{{\rm w},i,n}$, $M_{{\rm w},i,n}$, and $f_{{\rm w},i,n}$ for each SNe of the $n$-th halo are given by eq.~(\ref{energyout}), (\ref{massout}), and (\ref{metalout}), respectively, and $Y_{A,i,n}$ is the latter yield. 
}
The last terms in eqs. (\ref{windmass})-(\ref{windabund}) are the infall of the \red{wind matter into other mini-halos, located} in the metal enriched area by the wind.  
\red{When the wind from the $n$-th mini-halo encounters another ($m$-th) mini-halo, a part of the wind gas and metal are accreted onto the $m$-th mini-halo. 
We set the accretion rate of wind mass as gas infall rate, $\dot M_{{\rm in},m}$, of $m$-th proto-galaxy. 
Here $S_n$ is a set of mini-halos in the area enriched by the wind ejected from the $n$-th halo, and $\dot M_{{\rm in},m}$ is the infall rate to the $m$-th halos. 
W}e assume spatially random distribution of mini-halos and take the probability that a halo is in the region polluted by an $n$-th halo to be ${\cal M}_{n}/M_{\rm IGM}$ (see \S\ref{XigmS}). 
A mini-halo is surrounded by its own wind (i.e., $n \in S_n$), and a part of the wind matter returns to the mini-halo as it grows in mass by infall. 

When two mini-halos merge, the winds, which are ejected from the mini-halos, also ``merge''. 
The merger of winds is given by the pairwise sums of mass, momentum, and volume ($\equiv 4\pi {\cal R}^3/3$). 
The chemical abundance of winds is averaged. 
Until merger, each wind evolves independently even though their metal enriched areas may overlap. 

\subsubsection{Pre-Enrichment of Proto-Galaxies}\label{XigmS}
\red{We set the initial abundance of proto-galaxies formed with the pre-enriched IGM as follows. }

\red{A newly formed halo is pre-enriched by the wind from an $n$-th halo with probability of ${\cal M}_{n}/M_{\rm IGM}$. 
We simply assume the spatially random distribution of mini-halos. 
When an $m$-th halo is newly formed, we randomly set whether or not the halo is in the pre-enriched region by the wind ejected from the previously existing $n$-th halos, i.e., we randomly define $m \in S_n$ or not with the probability of ${\cal M}_{n}/M_{\rm IGM}$ by $m-1$ times. }

\red{
The initial abundance of a proto-galaxy is given by the sum of the elemental abundances of winds which cover the position where the proto-galaxy formed,} 
\begin{equation} \label{infallabund}
X_{{\rm IGM},A,m} = \sum_{\{n|m \in S_n\}} {\cal X}_{A,n}.
\end{equation}

\red{
The inflow gas to the existent proto-galaxies is also polluted by the same way. 
We redefine $S_n$ following the change of probability ${\cal M}_{n}/M_{\rm IGM}$ as the mass ${\cal M}_{n}$ of each wind evolves. }

\section{Results}\label{resultS}

\begin{figure}
\includegraphics[width=\columnwidth]{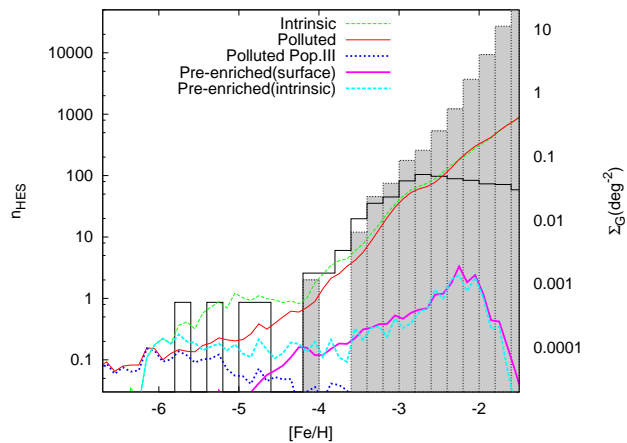}
\caption{Metallicity distribution functions (MDFs). 
Solid red line and dashed green line denote the predicted distributions of intrinsic metallicity ($\feoh_{\rm i}$) and after surface pollution ($\feoh_{\rm s}$), respectively. 
 Dotted blue line denotes that for Pop.~III survivors. 
The intrinsic and surface metallicity distributions for the first generation stars in the pre-enriched proto-galaxies are also plotted by dashed cyan and solid magenta lines, respectively. 
We compute the number of stars which are expected to be distributed in the survey volume of the HES survey (left axis) and the surface number density of giant stars in the MW halo (right axis). 
Histograms show the observational MDFs by the HES survey \citep[][gray shaded]{Schorck09} and the SAGA database (black solid line). 
The HES data is bias-corrected but disk stars can be contaminated at $\feoh > -2$. 
The SAGA sample is biased toward low-metallicity but is expected to be unbiased below $\feoh \lesssim -3$. 
The abundance determination is more accurate for the SAGA sample at $\feoh \lesssim -2.5$ because of higher resolution. 
For details about the observed data, see Section \ref{sampleS}.  
}
\label{MDF}
\end{figure}

Figure~\ref{MDF} shows the metallicity distribution function (MDF) for our model with the fiducial parameter set (Table~\ref{Tparam}). 
We show the predicted distribution of intrinsic metallicity (solid red line) and surface metallicity (dashed green line) of halo stars and compare them with the observed MDFs (histograms). 
The detailed descriptions about the observational data are in Section \ref{sampleS}. 
As shown in this figure, our model can reproduce not only the shape of the MDF but also the total number of the EMP survivors identified by the survey of the Milky Way halo. 


The accretion of ISM is the dominant source of iron for stars with $\feoh_{\rm s} \lesssim -4$. 
Most stars with $\feoh_{\rm s} \lesssim -5.5$ in this model are the surface-polluted Pop.~III survivors, as shown in the dotted blue line in Figure~\ref{MDF}. 
\red{This indicates that the observed HMP stars can be the polluted Pop.~III stars. 
Carbon-enhanced HMP stars are thought to suffer binary mass transfer in addition. 
The mass transfer from intermediate massive companions can reproduce the observed abundance anomalies of these stars \citep{Suda04, Nishimura09}. 
Recent discovery of stars with $\feoh = -4.89$ and without carbon enhancement \citep{Caffau12} indicates that the low-mass stars can be formed without atomic cooling by carbon or oxygen. 
}

The stars formed in mini-halos with $\Tv>10^4$K at redshift $z<20$ under the LW-background constitute $\sim 82\%$ of stars at $-4>\feoh>-6$ (UMP/HMP stars). 
The reason is as follows. 
Primordial proto-galaxies with small gas mass \red{($M_{\rm gas} \sim 10^5\msun$)} become $\feoh > -4$ \red{ by a single SN. 
Therefore, no UMP/HMP stars are formed as the second or later generations of stars. 
A few low-mass stars with $-6<\feoh_{\rm s}<-4$ are formed as polluted Pop.~III stars but the number of them is small due to the very high-mass IMF assumed for Pop.~III stars. }
In contrast, larger-mass proto-galaxies with $\Tv>10^4$K \red{($M_{\rm gas} \gtrsim 4 \times10^6\msun$)} remain to be in smaller metallicity of $\feoh<-4$ after a single CCSN. 
\red{The IMF in these proto galaxies are assumed to be the same with EMP stars and significant number of low-mass UMP/HMP survivors are formed as the second (or a few later) generation(s) of stars. }
Although we assume chemical homogeneity inside proto-galaxies, the elemental abundance in those larger proto-galaxies could be actually inhomogeneous. 
We will study about detailed abundance distributions in these large proto-galaxies and UMP/HMP stars in the forthcoming papers. 
On the other hand, most ($\sim88\%$) of the second generation stars with $-4<\feoh<-3$ are formed in proto-galaxies with $M_{\rm gas} < 10^6\msun$. 
In these small proto-galaxies, SN ejecta is thought to be well mixed into whole proto-galaxies before they cool to form the next generation of stars.  

In Figure~\ref{MDF}, we also show the distribution of the first generation stars which are formed in the pre-enriched proto-galaxies. 
The solid magenta and the dashed cyan lines are the distribution of these {\it pre-enriched first stars} after and before surface pollution, respectively. 
These pre-enriched first stars account for $\sim 2\%$ of EMP survivors and for $\sim 20\%$ of stars at $-4>\feoh>-5$. 

\begin{figure}
\plotone{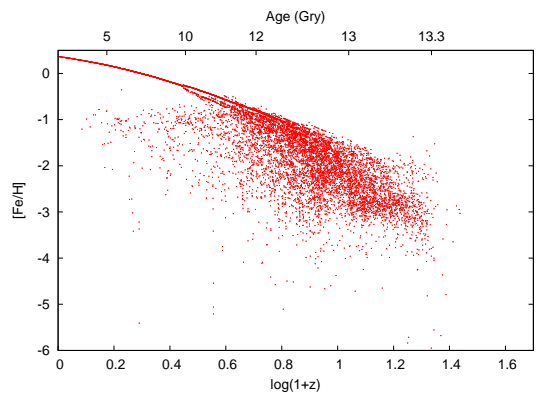}
\caption{The predicted distribution of surface metallicity against the formation redshift \red{(bottom axis) and stellar age (top axis)} for low-mass survivors. 
}\label{zFeH}
\end{figure}

As seen in Figure~\ref{zFeH}, the formation redshift of EMP stars is widely scattered. 
Pop.~III stars are formed around $z\sim 20$, and EMP stars mainly at $z\sim 15 \mhyph  5$. 
In the following subsections, we investigate the enrichment history of \rps element using this hierarchical chemical evolution model. 

\subsection{R-process Site}\label{MrS}

The dominant astronomical \rps site is not yet revealed as mentioned. 
In the fiducial model, we assign CCSNe as dominant \rps sources. 
Higher and lower mass limit ($M_{\rm hi}$ and $M_{\rm lo}$) for the production of \rps elements are free-parameters in this study. 
We summarize parameter values for \rps sites and yields of Ba and Eu in Table~\ref{Tmodel}.
We discuss NS merger as an \rps source in Section~\ref{NSMS}. 

We investigate the dependence on $M_{\rm hi}$ and $M_{\rm lo}$ by Models A-E.  
In Models A-E, we set mass range of \rps sites at the low-mass end of the CCSN mass range, following some previous chemical evolution studies \citep[e.g.][]{Mathews92, Ishimaru99, Qian08}. 
Stars around $\sim10\msun$ are thought to be explode as electron capture SNe (ECSNe) with O-Ne-Mg cores. 
ECSN is a possible \rps site as mentioned above \red{ \citep[but see also][]{Wanajo11a} }.   
\citet{Pumo09} argue that the mass range of progenitor stars for ECSNe is $9-10\msun$ for metal poor stars. 
We discuss the result of Model C, in which $(M_{\rm lo}, M_{\rm hi})=(9\msun, 10\msun)$, in detail as a best-fit typical case in Section~\ref{scenarioS} and parameter dependence in Section~\ref{paramS}. 

For simplicity, the mass of the ejected \rps elements per SN event is assumed to be constant over the mass range of each case. 
We determine the $\Y{Eu}$ to give $\langle\abra{Eu}{Fe}\rangle = 0.6$, where $\langle\abra{Eu}{Fe}\rangle$ is the IMF weighted average of relative abundance of ejecta of all CCSNe, and 0.6 is the averaged observational abundance of Population~II stars. 
Under this assumption, $\Y{Eu}$ is anti-proportional to the number fraction, $P$, of the \rps sources among CCSNe that produce iron: 
\begin{equation}
P \equiv\int^{M_{\rm hi}}_{M_{\rm lo}} \xi(m) dm \bigg/ \int^{40}_9 \xi(m) dm .
\label{eq:number_fraction}
\end{equation} 
   The yield of Ba is determined to be $\Y{Ba}/\Y{Eu}=7.6$ to give the same ratio as the solar \rps abundance ratio \citep{Arlandini99}.  

We also adopt some other values for $M_{\rm hi}, M_{\rm lo}$ and $Y$ for comparison in Section~\ref{paramS}. 
We compute models with the same mass ranges and elemental yields as \citet[][$(M_{\rm lo}, M_{\rm hi})=(20\msun,25\msun)$]{Argast04} or \citet[][$(M_{\rm lo}, M_{\rm hi})=(12\msun,30\msun)$]{Cescutti06}. 
They advocate mass dependent yields for Eu and Ba. 
The results are described in Section~\ref{argastS}.

\subsection{Basic Properties of the Hierarchical Chemical Evolution Model}\label{scenarioS}

\begin{figure}
\plotone{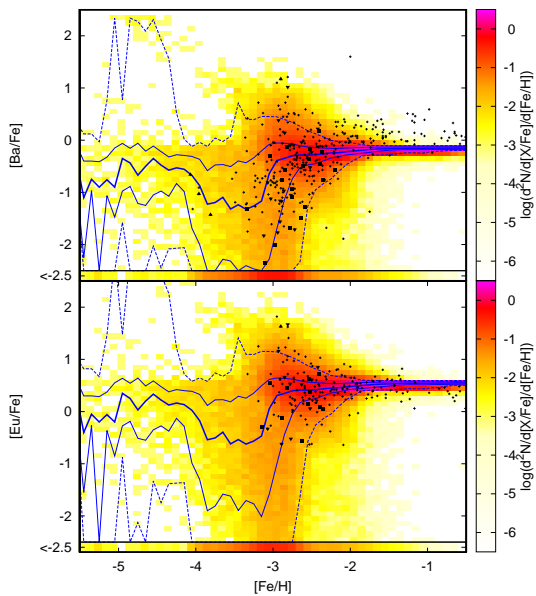}
\caption{The predicted distributions of Ba (top panel) and Eu (bottom panel) by Model C against the metallicity for the low-mass stars that survive to data. 
ECSNe of progenitor stars with $9$-$10 \msun$ are assumed to be the \rps sites. 
The predicted number density of stars per unit area of $\Delta \feoh = \Delta \abra{\it r}{Fe} = 0.1$ is color-coded. 
The blue lines are 5, 25, 50, 75, and 95 percentile curves of the predicted distributions. 
The observed data are shown by black symbols (see \S~\ref{sampleS} for details on the observational sample). 
}
\label{fiducial}
\end{figure}

\begin{figure}
\plotone{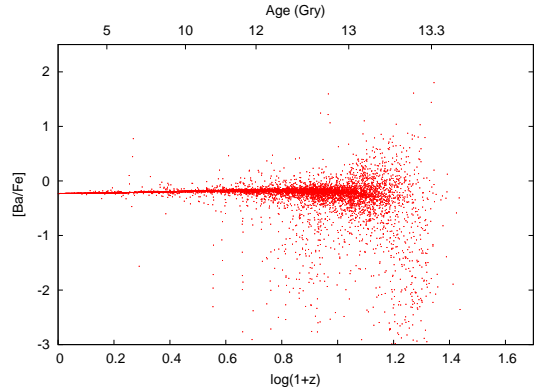}
\caption{The predicted distribution of Ba abundance of low-mass survivors against the formation redshift \red{(bottom axis) and stellar age (top axis)} from Model C. 
}\label{zBaFe}
\end{figure}

\begin{figure}
\plotone{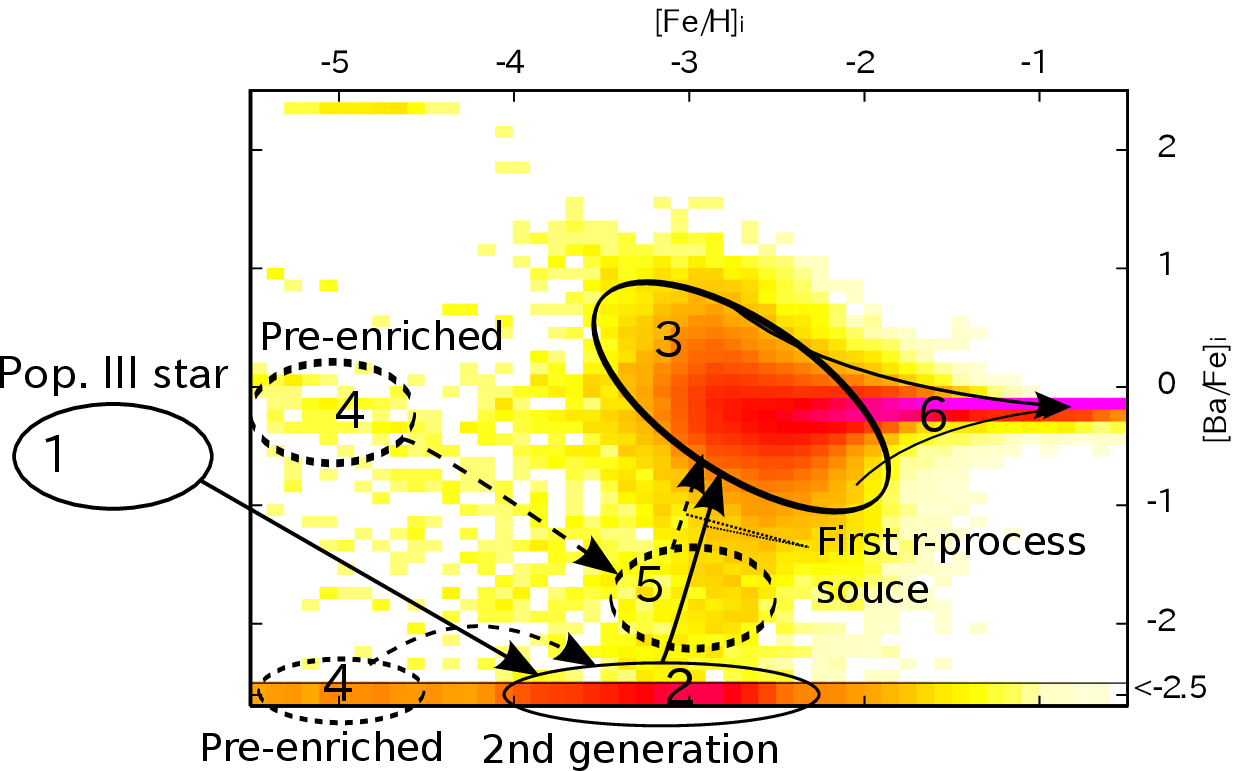}
\plotone{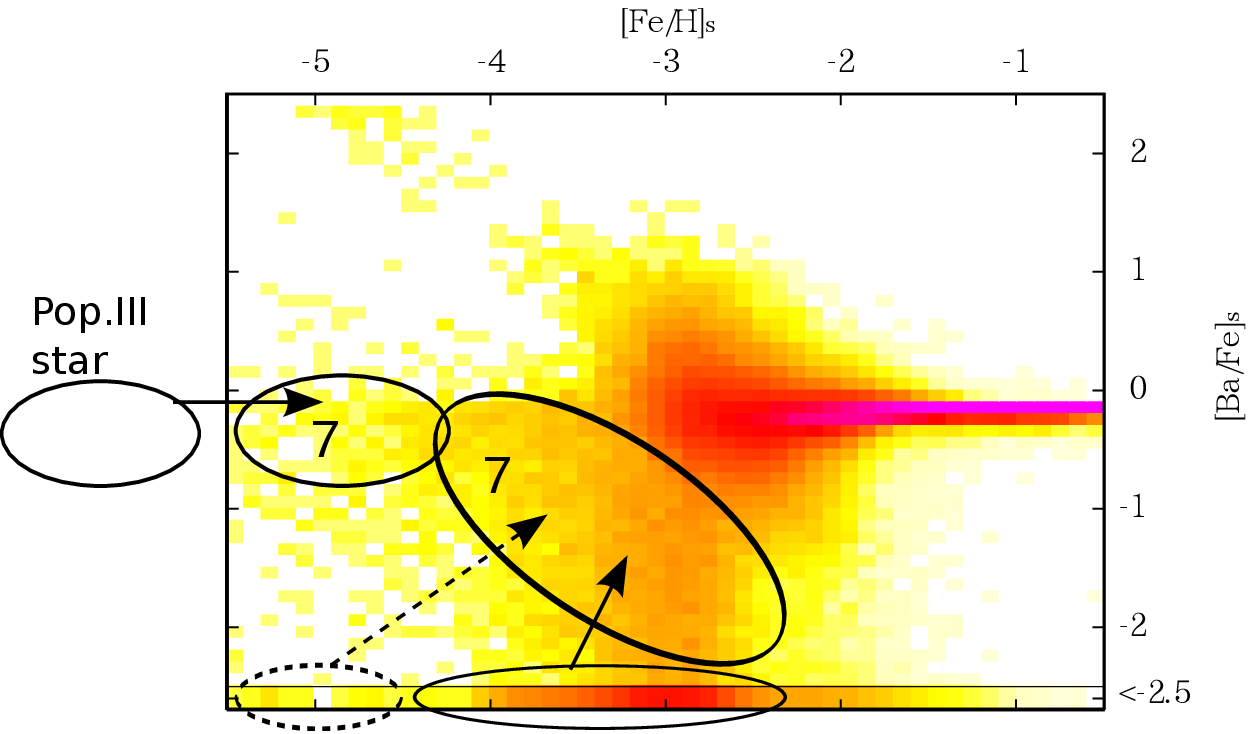}
\caption{
A schematic picture of the evolution process in our hierarchical chemical evolution model on the $\feoh$-$\abra{Ba}{Fe}$ plane. 
The color maps shows predicted number distribution of stars before (top panel) and after (bottom panel) the ISM accretion.  
The solid and dashed arrows show the typical evolutionary tracks of pristine and pre-enriched proto-galaxies, respectively. 
The numbers in the figure correspond to the itemized figures in the text (\S\ref{scenarioS}).  
Chemical abundance evolves as follows: 
At first, a PISN or a FeCCSN provides iron without providing \rps elements.  
A few tens of million years later, first ECSN makes its host proto-galaxy to $\abra{Ba}{H}\sim -3$ at $\feoh\sim -2$ to $-3.5$. 
As the chemical evolution progresses with subsequent SNe, $\abra{Ba}{Fe}$ is averaged owing to mixing ejecta of various SNe and as well as galaxy merger. 
On top of this, the surface pollution changes the the surface abundance and Pop.~III and \nor stars come to have $\abra{Ba}{H}\sim -5$. 
See text for details. 
}
\label{illust}
\end{figure}

In this subsection, we describe results of our fiducial model, Model C ($(M_{\rm lo}, M_{\rm hi})=(9\msun, 10\msun)$ \red{with $P=10\%$, $\Y{Ba} = 6.55\times 10^{-6}$, and $\Y{Eu} = 8.62\times 10^{-7}$} ), in detail as a typical case of our hierarchical model. 

Figure~\ref{fiducial} shows the predicted distribution of stars in the $\feoh$-$\abra{Ba}{Fe}$ (top panel) plane and the $\feoh$-$\abra{Eu}{Fe}$ plane (bottom panel). 
The color maps denote the predicted number density of giant stars.  
The number of stars with $\abra{X}{Fe}<-2.5$ in each $\feoh$ bin is color coded in the bottom cells of the panels. 
For EMP stars, \rps elements on such very \rps deficient stars are not detected by spectroscopic observations. 
The blue lines denote medians ({\it thick solid}), upper and lower quartiles ({\it thin solid}), and 5 and 95 percentiles ({\it dotted}) of the predicted distributions. 
We use the observational sample compiled by the Stellar Abundance for Galactic Archaeology (SAGA) database \citep[][black symbols]{Suda08}. 
Detailed description of the observational sample is in Section~\ref{sampleS}. 
We note that the SAGA sample is strongly biased toward lower metallicity. 
At $-2.8<\feoh<-1$, the metallicity distribution of the \red{SAGA sample} is almost flat as seen in Figure~\ref{MDF}. 
The predicted number density in Fig.~\ref{fiducial} and following figures of abundance distribution is scaled considering the bias at $\feoh>-2.8$. 
At $\feoh\le-2.8$, we plot the computation result directly because the SAGA sample is almost unbiased.  
In this study, since we set $\Y{Eu}/\Y{Ba}$ to be the solar \rps abundance ratio for all SNe, the predicted distribution of $\abra{Eu}{Fe}$ is the same as of $\abra{Ba}{Fe}$, but shifted by $\sim +0.7$ dex. 

As shown in Figure~\ref{zFeH} and Figure~\ref{zBaFe}, chemical evolution process is different from proto-galaxy to proto-galaxy. 
Abundance of the \rps elements of stars which are formed in small young proto-galaxies are scattered widely 
while stars around the averaged abundance are formed in the galaxies with larger mass and larger metallicity. 

Figure~\ref{illust} is a schematic picture of the chemical evolution process on the $\feoh$-$\abra{Ba}{Fe}$ plane in \red{the fiducial model}. 
Color in the top and bottom panels denote the predicted distributions of the intrinsic abundance and the polluted surface abundance of EMP survivors, respectively. 
In the top panel, the evolution of elemental abundances in typical proto-galaxies is illustrated by solid arrows and the evolution in the pre-enriched proto-galaxies is illustrated by dashed arrows. 
In the bottom panel, we illustrate the change of surface abundance by the ISM accretion. 
Elemental abundance evolves as follows: 

\begin{enumerate}
\item
In the beginning, a proto-galaxy is formed without iron and without \rps elements. 
Pop.~III stars are formed in small proto-galaxies. 
A few low-mass Pop.~III survivors are formed as secondary companions of Pop.~III binaries. 
Accretion of ISM changes their surface abundance and we observe the {\it polluted Pop.~III stars} at $\feoh\sim -5$. 

\item 
Progenitor mass of the first SN in the proto-galaxy is almost always larger than $10\msun$ because of the very high-mass IMF for Pop.~III stars and longer lifetime of stars with $9-10\msun$ than more massive stars. 
Therefore, the first SN yields iron but no \rps elements. 
Its host proto-galaxy becomes $\feoh \gtrsim -4$, but $X_{\rm Ba}$ remains at zero ($\abra{Ba}{H}=-\infty$). 
Subsequent FeCCSN(e) enrich host proto-galaxy to $-4 \lesssim \feoh \lesssim -2$. 
{\it No-r EMP stars} are formed here. 

\item
A few tens million years after the first SN, the first ECSN explodes and provides \rps elements. 
By a single ECSN, the host proto-galaxy becomes $-3.5 \lesssim \abra{Ba}{H} \lesssim -2$. 
Barium abundance relative to iron jumps from $\abra{Ba}{Fe}=-\infty$ to $\abra{Ba}{Fe}\gtrsim -1$. 
Diversity of mass of proto-galaxies and of iron abundances makes large scatter of the $\abra{\it r}{Fe}$ after this first \rps element injection. 
The short axis of the tilted ellipse ($\abra{\it r}{H}$) depends on gas mass and the long axis (along constant $\abra{\it r}{H}$) depends on the delay time of \rsource and the iron yield of previous SNe. 
When ECSN occurs at an early age in proto-galaxies with small gas mass, proto-galaxies become $\abra{Eu}{Fe} > +1$ and {\it r-II stars} are formed.

\item
Galactic winds with metal are ejected from proto-galaxies by energetic SNe and enrich the IGM with metal. 
The ejected matter is mixed and diluted with the IGM and $\feoh$ decreases. 
But $\abra{Ba}{Fe}$ does not change because both iron and barium are diluted in the same way. 
The typical evolutionary tracks of pre-enriched proto-galaxies are illustrated by dashed arrows and dashed circles in Fig.~\ref{illust}. 
Approximately $\sim4\%$ of proto-galaxies are formed with the pre-enrichmed IGM. 
{\it Pre-enriched first stars} are formed in these proto-galaxies. 
These stars are divided into two groups different in $\abra{Ba}{Fe}$.   
One comprises stars without Ba that are only polluted by PISNe and/or FeCCSNe. 
The other group comprises stars which contain \rps elements ejected from ECSNe. 
Initial Ba abundances of these stars are distributed around $\abra{Ba}{Fe}\sim -1$ to $0$. 

\item
Similar to the pristine proto-galaxies, the chemical abundance of pre-enriched proto-galaxies evolves by subsequent SNe.  
By FeCCSNe, the proto-galaxies become $ -4 \lesssim \abra{Fe}{H} \lesssim -2$. 
For proto-galaxies which are pre-enriched with \rps elements, $\abra{Ba}{Fe}$ decreases. 
When the first ECSNe explode, their barium abundances jump to $\abra{Ba}{Fe}\gtrsim -1$. 

\item
After the first ECSN, $\abra{Ba}{Fe}$ approaches the IMF weighted average of SN yields owing to mixture of ejecta from many SNe. 
Mergers of proto-galaxies also average their abundances. 
Scatter of the element abundances decreases as metallicity increases. 
At $\feoh\gtrsim-1.5$, the scatter is smaller than the typical uncertainty of the abundance measurement ($\sim0.2$dex). 

\item
Distribution of initial abundance shown in the top panel of Figure~\ref{illust} is given as described above. 
On top of this, the abundance distribution is changed by the surface pollution as shown in the bottom panel. 
Pop.~III survivors are polluted with metal and becomes UMP/HMP stars as seen in Figure~\ref{MDF}. 
No-{\it r} EMP stars are polluted with \rps element to $\abra{Ba}{H}\sim -5$ on average (tilted ellipse in the bottom panel of Fig.~\ref{illust}). 
We describe the effect of the ISM accretion in Section~\ref{accretionS} in detail.
\end{enumerate}

\red{ 
}

\subsubsection{Pre-Enrichment}\label{preenrichS}

Outflow from proto-galaxies forms galactic wind and enrich the IGM with metal. 
In our fiducial model, $\sim4\%$ of proto-galaxies in number is formed in the pre-enriched IGM. 
First generation stars in these pre-enriched proto-galaxies are expected to preserve the chemical signature of the outflow. 
As seen in Figure~\ref{MDF}, these stars distribute over wide metallicity range, $-6\lesssim\feoh_{\rm i}<-2$. 

A very massive first star explodes as PISN, and eject a large amount of metal, and form a galactic wind to enrich the ambient IGM with $\feoh\gtrsim-2$. 
A FeCCSN form a galactic wind with metallicity $\feoh\sim-4$ to $-3$. 
Winds can be diluted to $\feoh\sim-6$ as it spread in the IGM with $\sim10^{7-8}\msun$. 
As galaxies evolve, many SNe add metal and momentum to winds, and winds evolve. 
At $z= 20$ and $10$, $\sim 3\%$ and $34\%$, respectively, of the Galactic IGM is polluted with metal. 
$\sim 61\%$ of pre-enrichmed first stars have \rps elements at their birth. 
Barium abundance of these stars is $\abra{Ba}{Fe}_{\rm s} \sim -0.3$ on average. 

\begin{figure}
\plotone{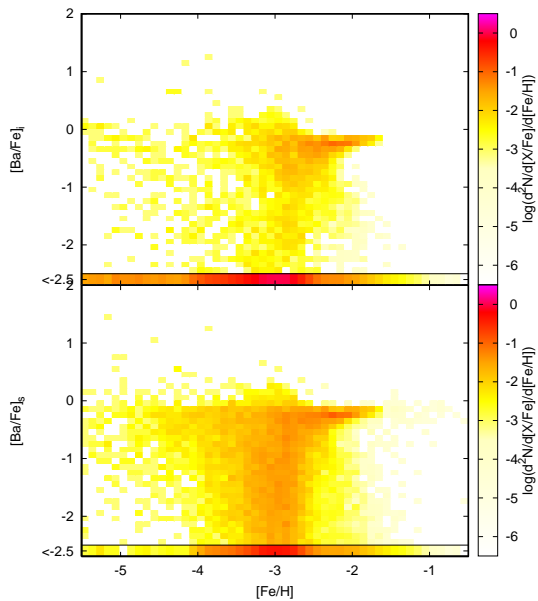}
\caption{ Distribution of EMP survivors which are formed before the first ECSN in their host halo. 
Top and bottom panels show the distributions before and after they suffer the accretion of ISM, respectively.  
The barium abundance of the top panel shows the contribution of pre-enrichment. 
}
\label{preenrich}
\end{figure}

Figure~\ref{preenrich} is abundance distribution of stars that they are formed before the first ECSNe in their host proto-galaxies. 
Distribution of their intrinsic abundance of barium (top panel) shows contributions of pre-enrichment to the barium abundance.  
There formed $\sim 27\%$ of EMP survivors before the first ECSN, and $\sim 29\%$ of them have \rps elements in their interior due to pre-enrichment. 
For the initial barium abundance of these stars, the median is $\abra{Ba}{Fe}_{\rm i} = -0.56$.   

The bottom panel shows the distribution with the surface pollution taken into account. 
This shows that the abundance distribution of pre-enriched EMP stars are changed and dominated by the ISM accretion for $\abra{Ba}{H} \lesssim -3.5$. 

\subsubsection{ISM Accretion}\label{accretionS}

Because only a small portion of IGM is polluted with metal by winds at high redshift, most proto-galaxies are formed with the primordial abundance. 
In our fiducial model, $\sim 1100$ low-mass Pop.~III stars are formed and survive to date in the whole Milky Way halo. 
The accretion of ISM determines the present surface element abundances of these Pop.~III survivors. 
In Paper~I, we computed the accretion rate of metal from ISM onto EMP survivors with both the chemical evolution and the merging history taken into account. 
We showed that Pop.~III survivors are polluted with $\feoh \sim -6$ and observed as UMP/HMP stars. 
In this model, the majority of HMP stars ($\feoh < -5$) are polluted Pop.~III survivors.
For these stars, the ISM accretion is only the source of metal on their surfaces.  

\begin{figure}
\plotone{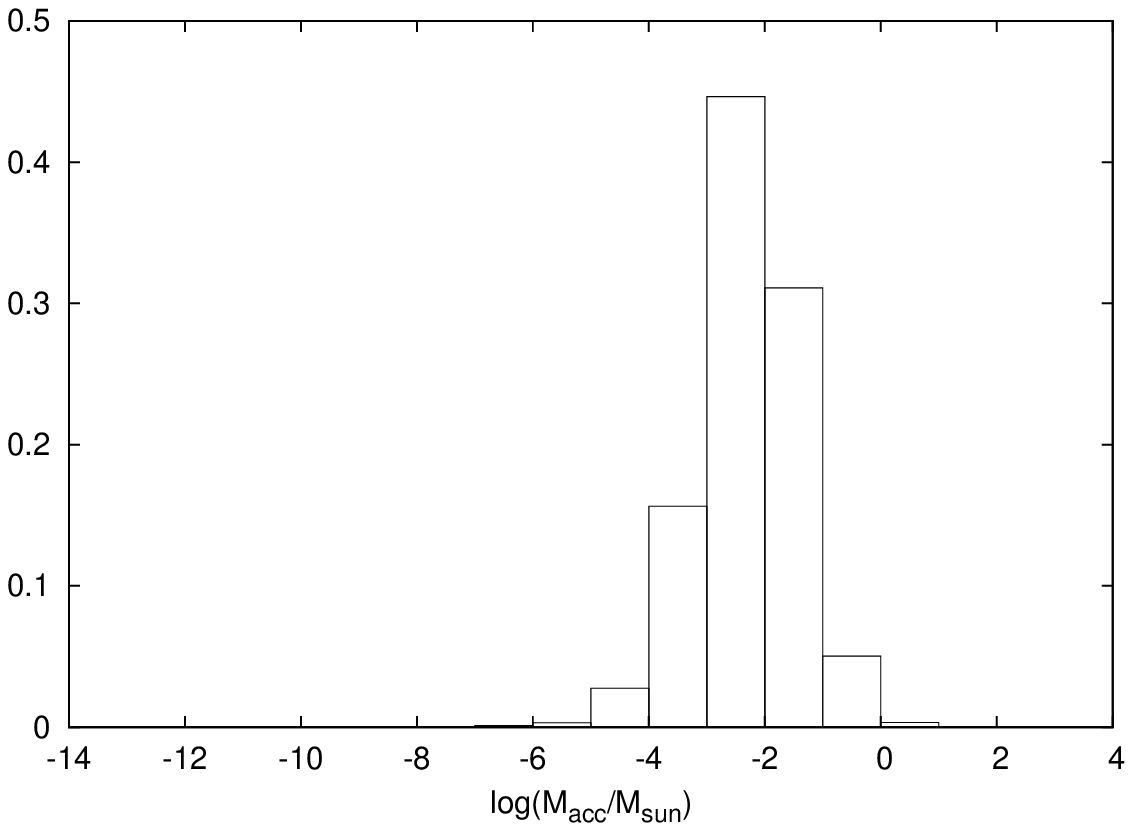}
\plotone{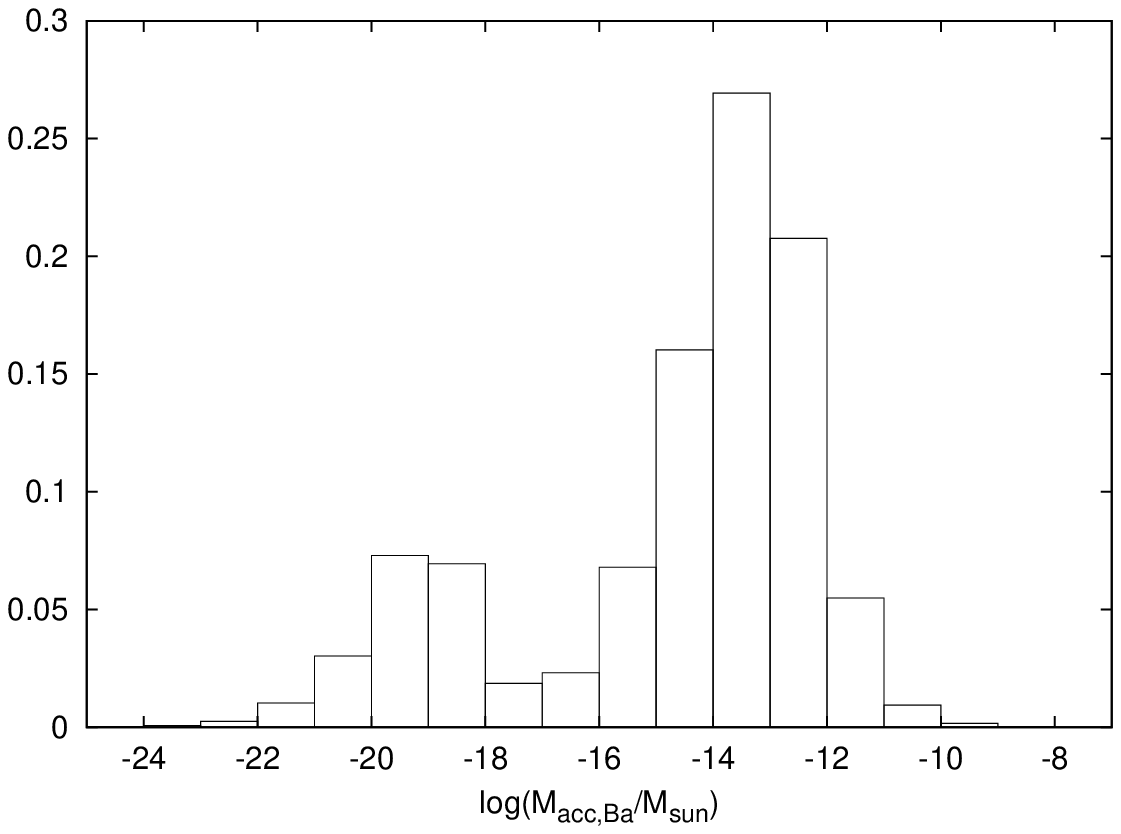}
\caption{
Distribution of the mass of the accreted ISM ({\it top panel}) and of the accreted barium ({\it bottom panel}) for \nor stars (stars formed without barium). 
The Y axis is the number fraction of stars. 
On average, surface abundance of giant stars are changed from $\abra{Ba}{H}=-\infty$ to $\abra{Ba}{H}=-5.4$ due to pollution by accretion of ISM. 
Pollution rate have large scatter with $\sim 10$ dex. 
See text for details. 
}
\label{BaAcc}
\end{figure}

A majority of EMP survivors with $\feoh<-3$ have no \rps elements at their formation although they have metal. 
The accretion of ISM is an exclusive source of Ba for these stars. 
Figure~\ref{BaAcc} denotes distributions of the mass, $M_{\rm acc}$, of the accreted ISM ({\it top panel}) and the mass, $M_{\rm acc,Ba}$, of the accreted barium ({\it bottom panel}) for low-mass survivors without barium at their birth, i.e., Pop.~III survivors, pre-enriched first stars without pristine \rps elements, and \nor stars. 
As described in Sec.\ref{polmodelS}, we follow the change in the surface abundance of each individual EMP survivor by the accretion of ISM and compute the distribution of \rps element abundances. 
The amount of accreted barium onto EMP stars is distributed over 10 dex depending on the merging histories and the chemical enrichment histories of their host halos.

The distribution of $M_{\rm acc,Ba}$ is bimodal. 
The lower and higher peaks correspond to proto-galaxies that merge to larger halos before and after the first ECSNe, respectively. 
As shown in our previous paper, the ISM accretion is efficient in the proto-galaxies in which EMP stars are formed ($\dot M_{\rm acc}\sim 10^{-11}\msun/$yr), but the accretion rate decreases sharply after the merging event since the relative velocity between stars and ISM increases (see Paper I for details). 
Stars formed in a proto-galaxy in which an ECSN took place before the first merging event are distributed at $M_{\rm acc,Ba}\sim 10^{-13}\msun$.  
When their host proto-galaxies merge before the first ECSNe, on the other hand, stars can accumulate a tiny amount of barium, $M_{\rm acc,Ba}\sim 10^{-19}\msun$, from ISM.  
The surface barium abundance becomes $\abra{Ba}{H}\sim -5$ for a typical giant star but is distributed between $\abra{Ba}{H}\sim -13$ to $-3$.

\begin{figure}
\plotone{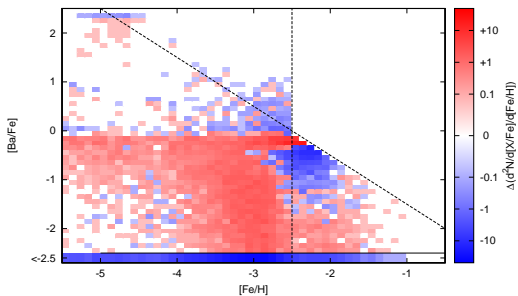}
\caption{
Change of the abundance distribution by the surface pollution on the $\feoh \abra{Ba}{Fe}$ plane. 
Differences in the number density of EMP giant survivors between the distributions of intrinsic and polluted surface abundances are color coded where red and blue mean the increase and decrease of number density by the surface pollution, respectively. 
We compute the changes of surface abundance only for stars with $\feoh<-2.5$ or $\abra{Ba}{H}<-2.5$. 
Blue color at the bottom cell denotes the number of these stars. 
In this figure, the number density is not scaled to match the observed number. 
}
\label{accretion}
\end{figure}

Figure~\ref{accretion} shows the effect of surface pollution by the accretion of ISM on the abundance distribution of EMP survivors. 
Difference in number density of EMP survivors between distributions of intrinsic abundance and polluted surface abundance are color coded. 
We compute the changes of surface abundance only for stars with $\feoh<-2.5$ or $\abra{Ba}{H}<-2.5$. 
\red{Blue at the bottom end of the figure denote number of stars without Ba in their interior. }
Most of these \nor stars are the early generations of stars formed before the explosion of the first \rsource in their host proto-galaxies but neither Pop.~III stars nor first generation stars. 
  The accretion of ISM is the dominant source on the surface of \nor stars and changes their abundance to the red area at $\feoh\sim -3$ to $-4$ and $\abra{Ba}{Fe} < 0$ in Figure~\ref{accretion}. 
After the surface pollution, the median of the barium abundance at $\feoh= -3$ to $-4$ becomes $\abra{Ba}{Fe}\sim -1$. 
For the majority of stars with $\abra{Ba}{H} \lesssim -3.5$, the accreted barium overwhelms the intrinsic barium in their surfaces convection zone. 

For $\abra{Ba}{H} \gtrsim -3.5$, the ISM accretion is not the dominant source of barium, and yet, has an effect on abundance distribution. 
The number density of stars at $\abra{Ba}{Fe}\sim-0.2$ increases but decreases in the other area; i.e., the abundance scatter was decreased. 
   This can be explained as follows; 
Accreted matter is chemically more evolved than EMP stars since the accretion continues after the ISM grows more metal-rich than the EMP stars.   
As chemically more evolved, $\abra{Ba}{Fe}$ converges to the IMF average, and the accretion of ISM with the averaged abundance decreases scatter of $\abra{Ba}{Fe}$.

\subsubsection{Most Metal Deficient Stars}\label{HMPS}

The abundance distribution at the most metal-poor range is distinctive in evaluating the effect of surface pollution and IGM pre-enrichment. 

Around $\feoh \sim -3$, our fiducial model predicts a trend of $\abra{r}{Fe}$ decreasing as metallicity decreases on average. 
For $\feoh<-3.3$, however, the trend diminishes and in the range of $-4 <\feoh < -3.3$, the median of $\abra{Ba}{Fe}$ remains almost with $\sim -1$.  
It is because the predicted \red{\rps element abundances at this metallicity range are} dominated by the ISM accretion. 
A majority of stars in this metallicity range are \nor survivors. 
Their surfaces are polluted with the \rps elements to $\abra{Ba}{Fe}\sim -1$ in the accreted ISM.  
Scatter of $\abra{Ba}{Fe}$ is very large since the accretion rate have large variations as mentioned above. 

Although the number of observed stars is not sufficient to determine whether $\abra{Ba}{Fe}$ keeps decreasing or a plateau is reached, the distribution of the SAGA sample seems to be consistent with our prediction.  
The typical barium abundance predicted is close to the detection limits of current observations and there are some stars on which no barium is detected. 
For europium, there are few stars with detection at $\feoh<-3.3$, and the abundance trend is not yet known. 

For $\feoh<-4$, our model predicts that a majority of stars are distributed at the interval of $-1<\abra{Ba}{Fe}<0$. 
For these stars, the ISM accretion is the dominant source for iron as well as \rps elements. 
The median of $\abra{Ba}{Fe}$ is higher for UMP survivors with $\feoh<-4$ than at $-4<\feoh<-3.3$ since the accretion raises $\abra{Ba}{H}$ similarly to stars with $-4<\feoh<-3.3$ while $\feoh$ is smaller. 
The predicted scatter of $\abra{Ba}{Fe}$ is smaller than in $-4<\feoh<-3.3$ for both the iron and barium abundances are dominated by the accreted matter and the relative amounts of barium and iron are not dependent on the accretion rate.

A distinctive feature of the ECSN scenario for \rps sources in our model is that it predicts stars with $\abra{Eu}{Fe} > +2.5$ at $\feoh<-4$. 
When the first SN in a mini-halo is a ECSN, it eject a small amount of iron and a large amount of \rps elements to form these UMP stars with large \rps elements enhancement. 
\footnote{ However, \citet{Wanajo03} argue that ECSNs yield \rps elements only when their explosion energy is large ($E_{\rm SN} = 3.5\times10^{51}$erg), and those with large explosion energy eject a larger amount ($\sim0.02\msun$) of iron than the $Y_{\rm Fe}$ of ECSN assumed in this study. }

Observationally, for $\feoh<-4$, we have only 3 samples with barium detected in $-4.2<\feoh<-4$, for which, $\abra{Ba}{Fe}$ is close to the median line in Figure~\ref{fiducial}. 
For $\feoh<-4.2$, 4 stars including 2 HMP stars have so far been identified but neither barium nor europium has been detected in any of them. 
\red{The observed upper limits of Ba abundances of these stars ($\abra{Ba}{Fe}<+0.82$ for HE0107-5240 and $\abra{Ba}{Fe}<+1.46$ for HE1327-2326) indicate that they are not stars with large \rps element enhancement. 
As shown in Figure~\ref{MDF}, there is a high likelihood that such metal deficient stars are the polluted Pop.~III survivors. 
We note that these 2 HMP stars with large carbon-enhancement are thought to suffer the binary mass transfer from intermediate-mass primaries. 
It may suggest that we should consider the s-process contribution for their surface element abundances. 
}

The result of our hierarchical chemical evolution model is distinct from previous inhomogeneous chemical evolution models especially in $\feoh<-3.3$. 
\citet{Ishimaru04} also argued SNe at the low-mass end of the CCSN mass range as \rps sites. 
But they predicted a monolithic decreasing trend as metallicity decreases at $\feoh<-3$. 
This is because they assume that SN ejecta is well mixed throughout all of the ISM after the SN triggers the star formation. 
\citet{Argast04} predicted higher typical abundance and large scatter, because they neglected \nor stars and assumed the smaller mixing mass for SN ejecta. 
\citet{Cescutti08} predicted no stars below $\feoh<-3.5$. 
We consider the distributions of Pop.~III survivors and \nor stars with the ISM accretion taken into account, but these previous studies do not. 
Additionally, we consider the realistic merger trees and inhomogeneity of IGM. 
The effects of the merging history, the inhomogeneity of IGM, and the ISM accretion will be important to interpret future observations of \rps element abundances of most metal-poor stars.

\subsection{Parameter Dependence}\label{paramS}
\begin{table*}
\begin{center}
\caption{Models (\rps sites)}
\label{Tmodel}

\scalebox{0.7}{
\begin{tabular}{|l|c|r|cc|l|l|}
\hline
Model	& $M_{\rm lo}$--$M_{\rm hi}$($\msun$) \tablenotemark{a}&
 $P$\tablenotemark{b} & 
$\Y{Ba}(\msun)$ \tablenotemark{c} & $\Y{Eu}(\msun)$ \tablenotemark{d} &
 $\epsilon_\star$ (yr$^{-1}$) \tablenotemark{e}& note \\
\hline
A & $9-9.1$& $1.06\%$ & $6.07\times 10^{-5}$ & $7.99\times 10^{-6}$ & $10^{-10}$ &  \\
B & $9-9.3$& $3.14\%$ & $2.48\times 10^{-5}$ & $2.70\times 10^{-6}$ & $10^{-10}$ &  \\
C\tablenotemark{f}	& $9-10$	& $10.0\%$ & $6.55\times 10^{-6}$ & $8.62\times 10^{-7}$ & $10^{-10}$ &  \\ 
D & $9-13$& $34\%$  & $1.88\times 10^{-6}$ & $2.47\times 10^{-7}$ & $10^{-10}$ &  \\
E	& $9-40$	& $100\%$  & $6.58\times 10^{-7}$ & $8.66\times 10^{-8}$ & $10^{-10}$ &  \\
H	& $30-40$	& $10.1\%$  & $6.53\times 10^{-7}$ & $8.59\times 10^{-8}$ & $10^{-10}$ &  \\
& & & & & &\\
C01 & $9-10$	& $10.0\%$ & $6.55\times 10^{-6}$ & $8.62\times 10^{-7}$ & $10^{-11}$ &  \\
C03 & $9-10$	& $10.0\%$ & $6.55\times 10^{-6}$ & $8.62\times 10^{-7}$ & $3\times10^{-11}$ &  \\
C3 & $9-10$	& $10.0\%$ & $6.55\times 10^{-6}$ & $8.62\times 10^{-7}$ & $3\times10^{-10}$ &  \\
C10 & $9-10$	& $10.0\%$ & $6.55\times 10^{-6}$ & $8.62\times 10^{-7}$ & $1\times10^{-9}$ &  \\
S & $9-10$	& $15.3\%$ & $3.81\times 10^{-6}$ & $5.01\times 10^{-7}$ & $10^{-10}$ & \citet{Chabrier03} IMF \\
S0 & $9-9.62$	& $10.0\%$ & $6.55\times 10^{-6}$ & $8.62\times 10^{-7}$ & $4.82\times10^{-10}$ & \citet{Chabrier03} IMF \\
Argast	& $20-25$	& $11.7\%$ & \multicolumn{2}{c|}{\citet{Argast04}; Model SN2025 }   & $10^{-10}$ &  \\
Cescutti & $12-30$	& $62\%$ & \multicolumn{2}{c|}{\citet{Cescutti06}; Model 1}  & $10^{-10}$ &  \\
& & & & & &\\
NSM6		& $9-25$ (binary) & $11.2\%$ & $5.72\times 10^{-6}$ & $7.53\times 10^{-7}$ & $10^{-10}$ & 
$t_c \tablenotemark{g}=10^6$yr, $p_{\rm NSM} \tablenotemark{h}=10\%$, $\epsilon_{\rm NSM} \tablenotemark{i}=10^{50}$erg \\ 
NSM7		& $9-25$ (binary) & $11.2\%$ & $5.72\times 10^{-6}$ & $7.53\times 10^{-7}$ & $10^{-10}$ & $t_c=10^7$yr, $p_{\rm NSM}=100\%$, $\epsilon_{\rm NSM}=10^{50}$erg \\ 
NSM8		& $9-25$ (binary) & $11.2\%$ & $5.72\times 10^{-6}$ & $7.53\times 10^{-7}$ & $10^{-10}$ & $t_c=10^8$yr, $p_{\rm NSM}=100\%$, $\epsilon_{\rm NSM}=10^{50}$erg \\ 
NSM7s		& $9-25$ (binary) & $11.2\%$ & $5.72\times 10^{-6}$ & $7.53\times 10^{-7}$ & $10^{-11}$ & $t_c=10^7$yr, $p_{\rm NSM}=100\%$, $\epsilon_{\rm NSM}=10^{50}$erg \\ 
NSM7p10		& $9-25$ (binary) & $1.12\%$ & $5.72\times 10^{-5}$ & $7.53\times 10^{-6}$ & $10^{-10}$ & $t_c=10^7$yr, $p_{\rm NSM}=10\%$, $\epsilon_{\rm NSM}=10^{50}$erg \\ 
NSM7e		& $9-25$ (binary) & $11.2\%$ & $5.72\times 10^{-6}$ & $7.53\times 10^{-7}$ & $10^{-10}$ & $t_c=10^7$yr, $p_{\rm NSM}=100\%$, $\epsilon_{\rm NSM}=10^{51}$erg \\ 
\hline

\end{tabular}
}

\tablenotetext{a}{Higher and lower mass limit for \rps sites}
\tablenotetext{b}{Percentage of \rps sources among CCSNe}
\tablenotetext{c}{Ba yield}
\tablenotetext{d}{Eu yield}
\tablenotetext{e}{Star formation efficiency}
\tablenotetext{f}{Fiducial model}
\tablenotetext{g}{Coalescence timescale of neutron stars binary}
\tablenotetext{h}{Coalescence frequency of neutron star binary}
\tablenotetext{h}{Kinetic energy of ejecta from neutron star merger}
\end{center}
\end{table*}

\begin{figure*}
\includegraphics[width=\textwidth]{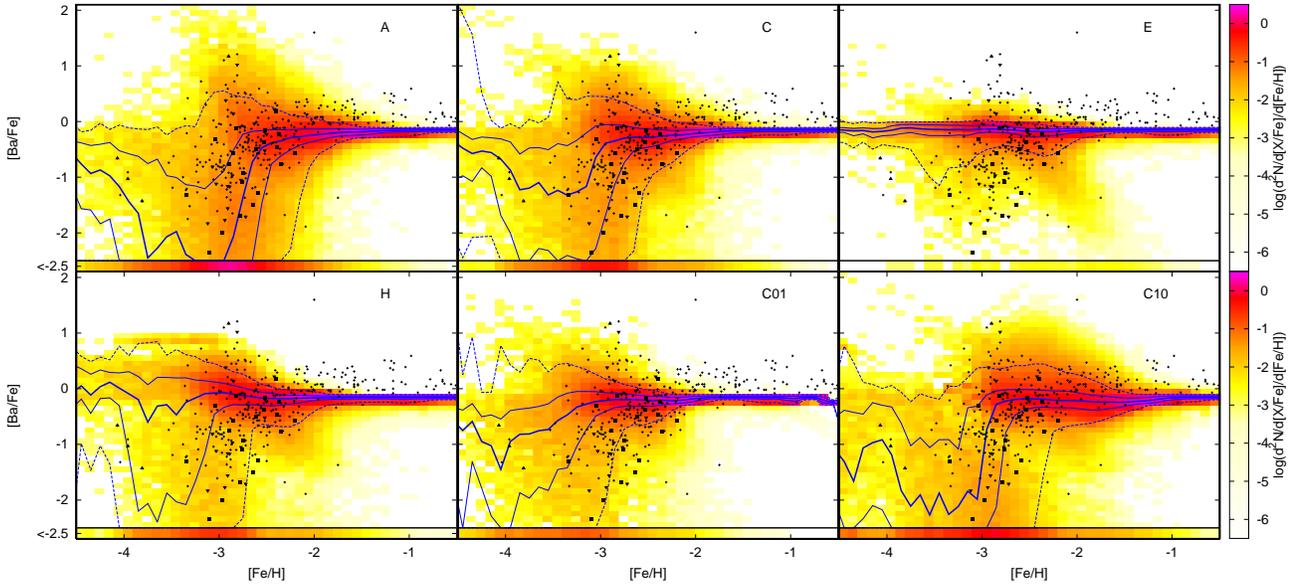}
\caption{The predicted abundance distributions of EMP survivors on the $\feoh$ and $\abra{Ba}{Fe}$ plane for the different parameter-sets on the \rps sites, Models A ({\it top left}), C ({\it top middle}), E ({\it top right}), and H ({\it bottom left}), and on the star formation rates, Models C01({\it bottom middle}) and C10({\it bottom right}). 
Color code, lines and symbols are the same with Fig.~\ref{fiducial}. 
}
\label{param}
\end{figure*}

\begin{figure*}
\includegraphics[width=\textwidth]{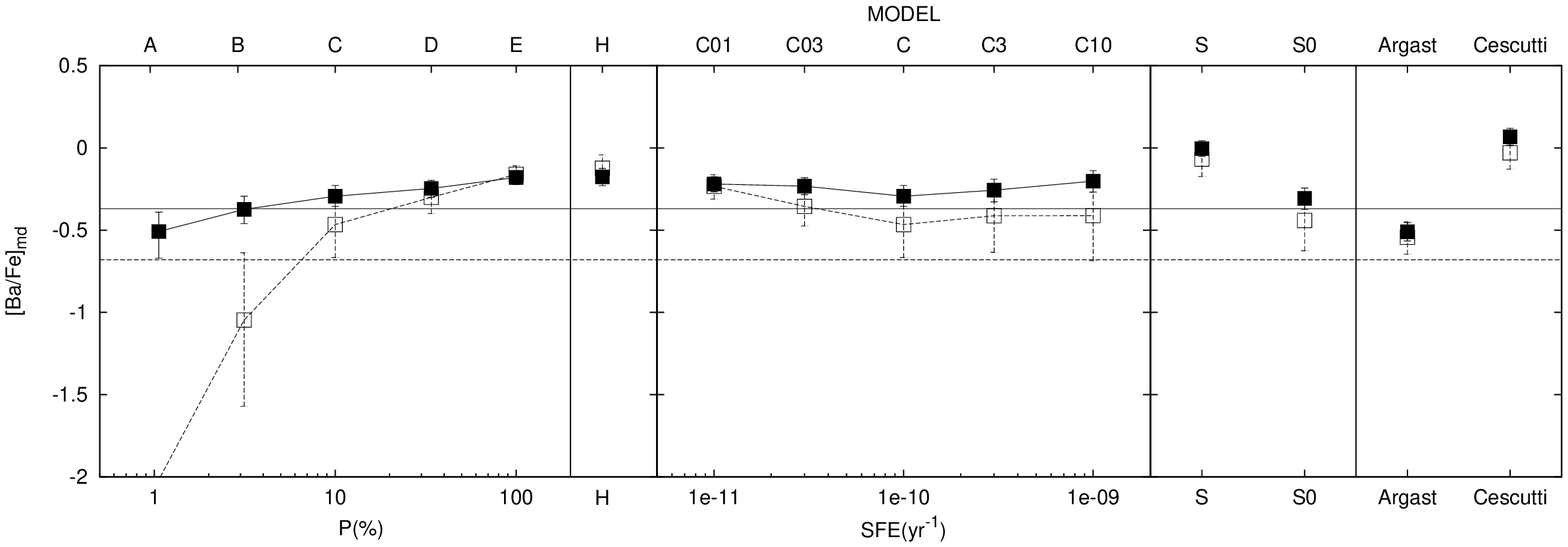} 
\includegraphics[width=\textwidth]{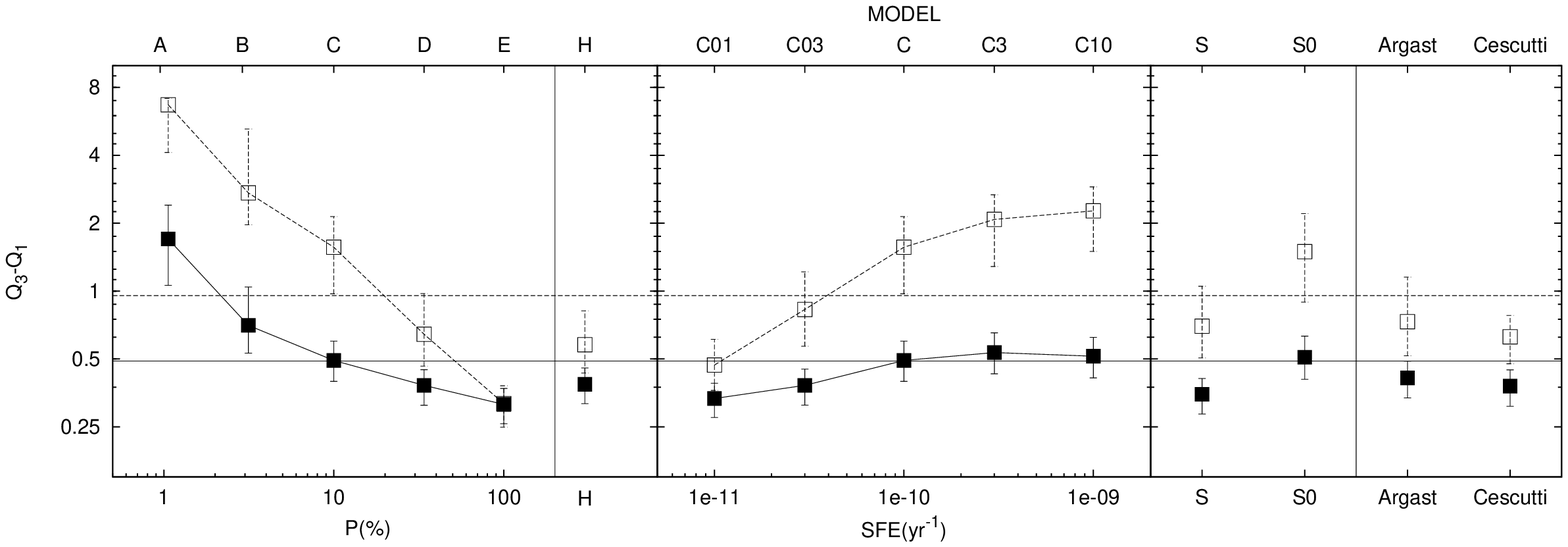} 
\includegraphics[width=\textwidth]{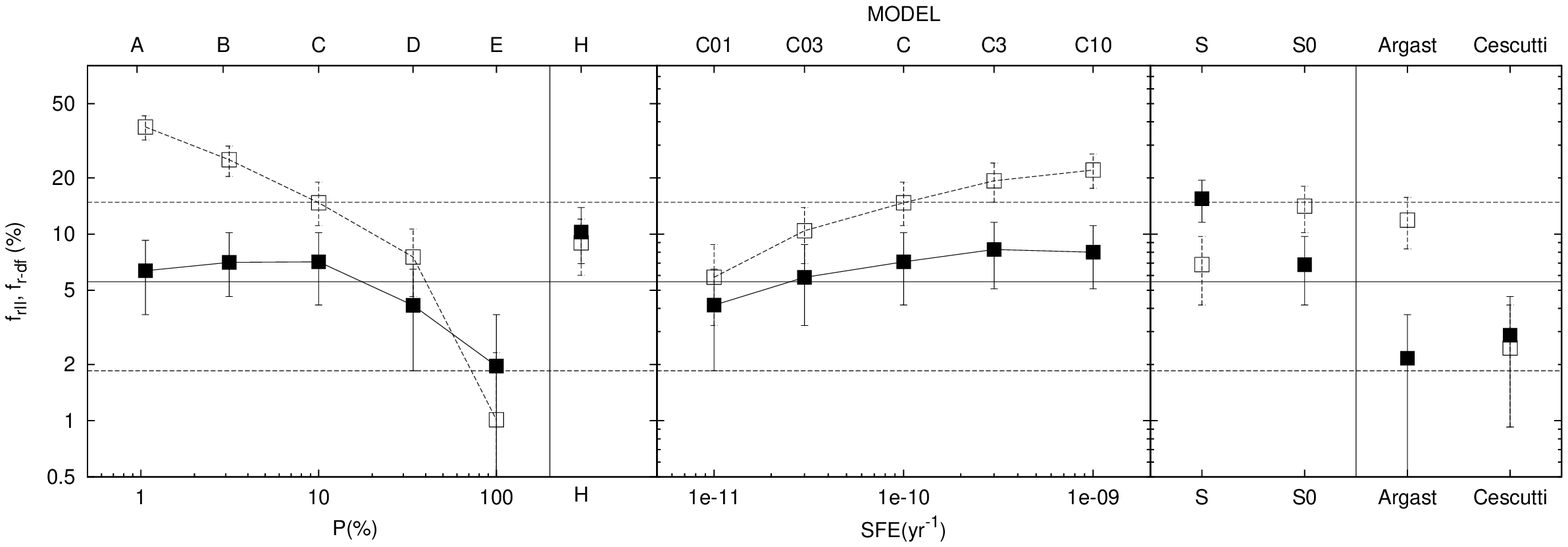} 
\caption{
Parameter dependence of the computation results and comparison with the observations. 
Models A--E show dependence on the width of the mass range of \rps source: 
C01--C10 show dependence on star formation rate: and 
S and S0 show results with low-mass IMF.  
Results with other scenarios for \rps sites are also presented (see Section~\ref{argastS}). 
Error bars are the $90\%$ bootstrap confidence intervals. 
Computation results are convolved with a Gaussian error in the observations of $\sigma = 0.2$ dex. 
[Top panel]: 
Median of the predicted abundance at $-2.8<\feoh\leqq-2.3$ ({\it filled square with solid lines}) and $-3.3<\feoh\leqq-2.8$ ({\it open square with dashed lines}), respectively. 
The horizontal straight lines show median of the observed abundance for $-2.8<\feoh\leqq-2.3$ ({\it solid}) and $-3.3<\feoh\leqq-2.8$ ({\it dashed}).
[Middle panel]:
Typical abundance scatter of $\abra{Ba}{Fe}$. 
We plot the interquartile ranges of barium abundance at metallicity ranges $-2.8<\feoh\leqq-2.3$ ({\it filled squares with solid lines}) and $-3.3<\feoh\leqq-2.8$ ({\it open squares with dashed lines}), respectively. 
The horizontal lines show the scatter of the observed samples at the $-2.8<\feoh\leqq-2.3$({\it solid}) and $-3.3<\feoh\leqq-2.8$({\it dashed}). 
[Bottom panel]
Number fraction of {\it r}-II stars ($\abra{Eu}{Fe}>+1$, {\it filled square with solid lines}) and \rdemp stars ($\abra{Ba}{Fe}<-5.5$, {\it open square with dashed lines}) among EMP survivors. 
The solid horizontal line is the observed number fraction of {\it r}-II stars. 
The lower dashed horizontal lines denote observed percentage of stars with report of upper limit of $\abra{Ba}{Fe}$. 
When we take into account for stars without report of $\abra{Ba}{Fe}$, the upper limit of $f_{r \rm df}$ can be up to the upper dotted horizontal line. 
}\label{allresult}
\end{figure*}

\red{We examine the dependence on the parameters of $M_{\rm hi}$, $M_{\rm lo}$, SFE, and IMF. 
Figure~\ref{param} shows the predicted distributions of stars in the $\feoh$-$\abra{Ba}{Fe}$ planes for 6 models. 
We summarize the parameter dependence of quantitative results in Table~\ref{Tresult} and in Figure~\ref{allresult}. }

\begin{table*}
\begin{center}
\caption{Model Results}
\label{Tresult}

\begin{tabular}{|l|cc|cc|ccc|}
\hline
&  &  & \multicolumn{2}{c|}{$Q_{3}-Q_{1}(\abra{Ba}{Fe})$ \tablenotemark{c}} &  &  & \\
Model	& $\abra{Ba}{Fe}_{\rm md}$ \tablenotemark{a} & $\Delta\abra{Ba}{Fe}_{\rm md}  \tablenotemark{b}$ &  
$(-2.8,-2.3]$ & $(-3.3,-2.8]$ & 
$f_{\rm rII}$ \tablenotemark{d}& 
$\feoh_{\rm rII}$  \tablenotemark{e}& 
$f_{r \rm df}$  \tablenotemark{f} \\
\hline
A		& -0.51 & -1.51 & 1.70 & 6.70 & 6.4\% & -2.84 & 38\% \\
B		& -0.37 & -0.67 & 0.70 & 2.73 & 7.1\% & -2.87 & 25\% \\
C		& -0.29 & -0.17 & 0.49 & 1.57 & 7.1\% & -2.90 & 15\% \\
D		& -0.25 & -0.05 & 0.38 & 0.64 & 4.2\% & -2.97 & 7.5\% \\
E		& -0.18 &  0.02 & 0.31 & 0.32 & 2.0\% & -2.96 & 1.0\% \\
H		& -0.18 &  0.05 & 0.39 & 0.58 & 10\% & -3.06 & 9.0\% \\
&  &  &  &  &  &  & \\
C01		& -0.22 & -0.01 & 0.33 & 0.47 & 4.2\% & -3.08 & 5.9\% \\
C03		& -0.23 & -0.12 & 0.38 & 0.83 & 5.9\% & -3.00 & 10\% \\
C3 		& -0.26 & -0.16 & 0.53 & 2.08 & 8.3\% & -2.81 & 19\% \\
C10 	& -0.20 & -0.21 & 0.51 & 2.27 & 8.0\% & -2.72 & 22\% \\
S		&  0.00 & -0.06 & 0.34 & 0.70 & 15\% & -2.89 & 6.9\% \\
S0		& -0.31 & -0.13 & 0.51 & 1.50 & 6.9\% & -2.89 & 14\% \\
&  &  & & & & &\\
Argast	& -0.51 & -0.03 & 0.41 & 0.73 & 2.2\% & -3.18 & 12\% \\
Cescutti&  0.07 & -0.10 & 0.38 & 0.62 & 2.9\% & -3.03 & 2.5\% \\
&  &  & & & & &\\
NSM6	& -0.20 & -0.09 & 0.40 & 0.83 & 6.0\% & -2.91 & 11\% \\
NSM7	& -0.28 & -0.14	& 0.47 & 1.37	& 6.9\% & -2.88	& 13\% \\
NSM8	& -0.64	& -0.92	& 1.07	& 2.95	& 3.4\% & -2.81	& 32\% \\
NSM7s	& -0.22	& -0.01	& 0.33	& 0.45	& 3.3\% & -3.03	& 5.4\% \\
NSM7p10	& -0.49	& -1.46	& 1.57	& 6.83	& 6.4\% & -2.84	& 36\% \\
NSM7e	& -0.31	& -0.14	& 0.46	& 1.48	& 3.7\% & -2.92	& 15\% \\
\hline
Observation & -0.37 & -0.31 & 0.5 & 1.0 & 12/216 & -2.83 & 4/216 -- 32/216 \\
&  &  & &  & (5.6\%) &  & (1.9\% -- 15\%) \\
\hline
\end{tabular}
\tablenotetext{a}{Median of $\abra{Ba}{Fe}$ at $-2.8<\feoh\le-2.3$}
\tablenotetext{b}{Difference between median of $\abra{Ba}{Fe}$ at $-2.8<\feoh\le-2.3$ and at $-3.3<\feoh\le-2.8$. Slope of the decreasing trend as metallicity decreases around $\feoh=-3$.} 
\tablenotetext{c}{Scatter of the abundance distribution. 
Interquartile range of $\abra{Ba}{Fe}$ at metallicity ranges of $-2.8<\feoh\le-2.3$ or $-3.3<\feoh\le-2.8$. }
\tablenotetext{d}{Percentage of {\it r}-II stars ($\abra{Eu}{Fe}>+1.0$) among giant EMP survivors}
\tablenotetext{e}{Average metallicity of {\it r}-II stars}
\tablenotetext{f}{Percentage of \rdemp stars ($\abra{Ba}{H}<-5.5$) among giant EMP survivors}
\end{center}
\end{table*}

\subsubsection{Mass Range of the R-process Site}\label{rsiteS}

The abundance trend and scatter are dependent on the mass range $(M_{\rm lo}, M_{\rm hi})$ of the \rps sites. 
The predicted typical abundances, trends, and scatters are given in Table~\ref{Tresult} and illustrated as a function of the model parameters in Figure~\ref{allresult}. 

Among the sample stars of the SAGA database, there is the trend of $\abra{Ba}{Fe}$ decreasing as the metallicity decreases.  
   Figure~\ref{param} shows that Models A and C predict a trend of decreasing $\abra{\it r}{Fe}$ due to long delay time to the explosion of ECSNe. 
On the other hand, Model H (($M_{\rm lo}, M_{\rm hi})=(30\msun,40\msun$) have the median of $\abra{Ba}{Fe}$ almost constant because of the short lifetime of the \rps sources. 

Scatter of the \rps element abundances depends on the number fraction, $P$, of \rps sources among the CCSNe in the whole mass range, as given in Table~\ref{Tmodel}. 
The scatter increases as $P$ decreases (see Figure~\ref{param}, Table~\ref{Tresult} and the middle panel of Figure~\ref{allresult}). 
Small $P$ means low event rate and large $\Y{Ba}$ and makes large scatter. 
To quantify the scatter, we use the interquartile range, $Q_{3}-Q_{1} (\abra{Ba}{Fe})$; the difference between the first and third quartiles of $\abra{Ba}{Fe}$ for $-2.8<\feoh\leqq-2.3$ ({\it filled square} in the middle panel of Figure~\ref{allresult}) and for $-3.3<\feoh\leqq-2.8$ ({\it open square}).  
The predicted values of $Q_{3}-Q_{1}$ in Table~\ref{Tresult} and Figure~\ref{allresult} are convolved with a Gaussian observational error of $\sigma = 0.2$ dex. 

The top right panel of Figure~\ref{param} shows the result of an extreme case with $P=100\%$, i.e., all CCSNe eject \rps elements (Model~E). 
In this case, the percentile curves of $\abra{Ba}{Fe}$ are almost flat against $\feoh$ and the predicted scatter is very small. 

 In the fiducial model, $\sim 4 \%$ of stars with $\feoh < -4$ show very large \rps element enhancement ($\abra{Ba}{Fe} \gtrsim +2$). 
They are the second generation stars born only with the ejecta of an ECSN. 
Since the iron yield of ECSN is much smaller than that of FeCCSN, they show very low metallicity and very large enhancement of \rps elements.  
The presence of these stars is a distinctive feature of the ECSN scenario for \rps site.

\subsubsection{The Star Formation Efficiency}\label{resultSFES}

The abundance distributions also depend on the star-formation efficiency (SFE). 
Model C01 with low SFE of $\epsilon_\star = 10^{-11}$/yr predicts a shallower slope of the median curve below $\abra{Ba}{Fe}=-3$. 
This is because that the lower SFE makes the progress of metal enrichment slower and the time-scale of chemical enrichment longer. 
Accordingly, ECSNe occur at lower metallicity, which shifts the downturn to lower metallicity and makes the declining slope gentle.   

The scatter of the abundances also decreases as SFE decreases since the proto-galaxies merge and their element abundances are averaged at lower metallicity owing to slower chemical evolution.  

On the other hand, Model C10 with high SFE predicts larger scatter and larger number of \rdemp stars. 

These results indicate that $\epsilon_\star$ should be $\sim10^{-10}$/yr to reproduce the observed abundance distribution. 

\begin{figure}
\plotone{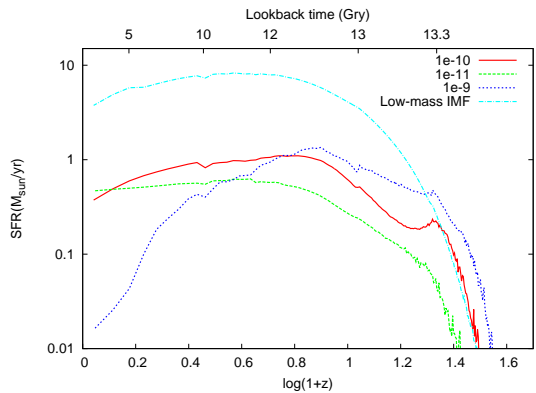}
\caption{Resultant total star formation rate history in the computations. 
The solid red, dashed green and dotted blue line describe results of Model C, C01, and C10, respectively. 
The dash dotted cyan line describes the result of Model S. 
See text for details. }
\label{SFR}
\end{figure}

 Figure~\ref{SFR} shows 
 the sum of the star formation rate for all proto-galaxies as a function of redshift. 
First stars are formed around $z \sim 30$, and SFR increases as new mini-halos collapse and accumulate gas. 
SFR once decreases at $z\geq 20$ due to the suppression of star formation by Lyman-Werner background, but increases again as galaxies grow in mass. 
In the fiducial model, SFR is peaked at $z\sim 5$ and decrease for lower redshift since gas in the galaxies are exhausted by star formation and SN driven outflow, although we have to take into account variations of IMF in such a low redshift.  
The resultant SFR is not proportional to SFE since supernova-driven outflow reduces gas mass in proto-galaxies and regulates star formation.

\subsubsection{The IMF of EMP Stars}\label{resultIMFS}

We use the high-mass IMF as a fiducial one according to our previous studies \citep{Komiya07, Komiya09}. 
In this section, we adopt the present-day IMF by \citet{Chabrier03} to discuss the IMF dependence of the predicted abundance distribution. 

\begin{figure}
\plotone{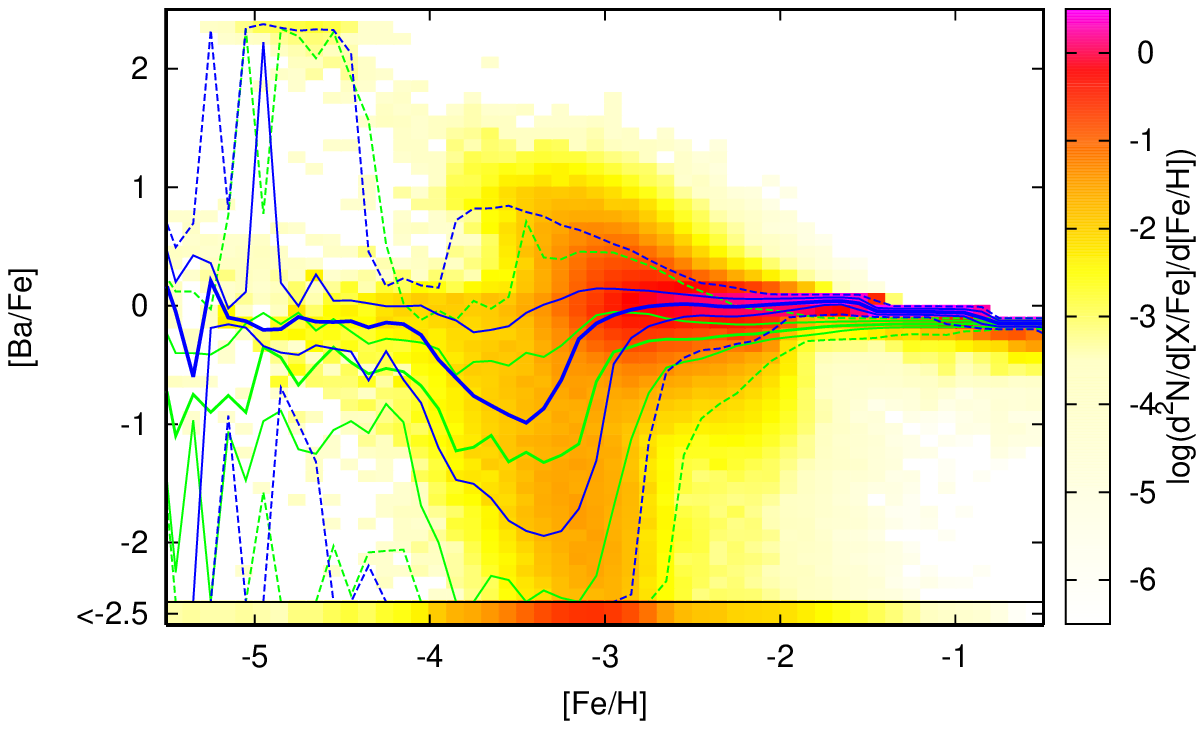}
\plotone{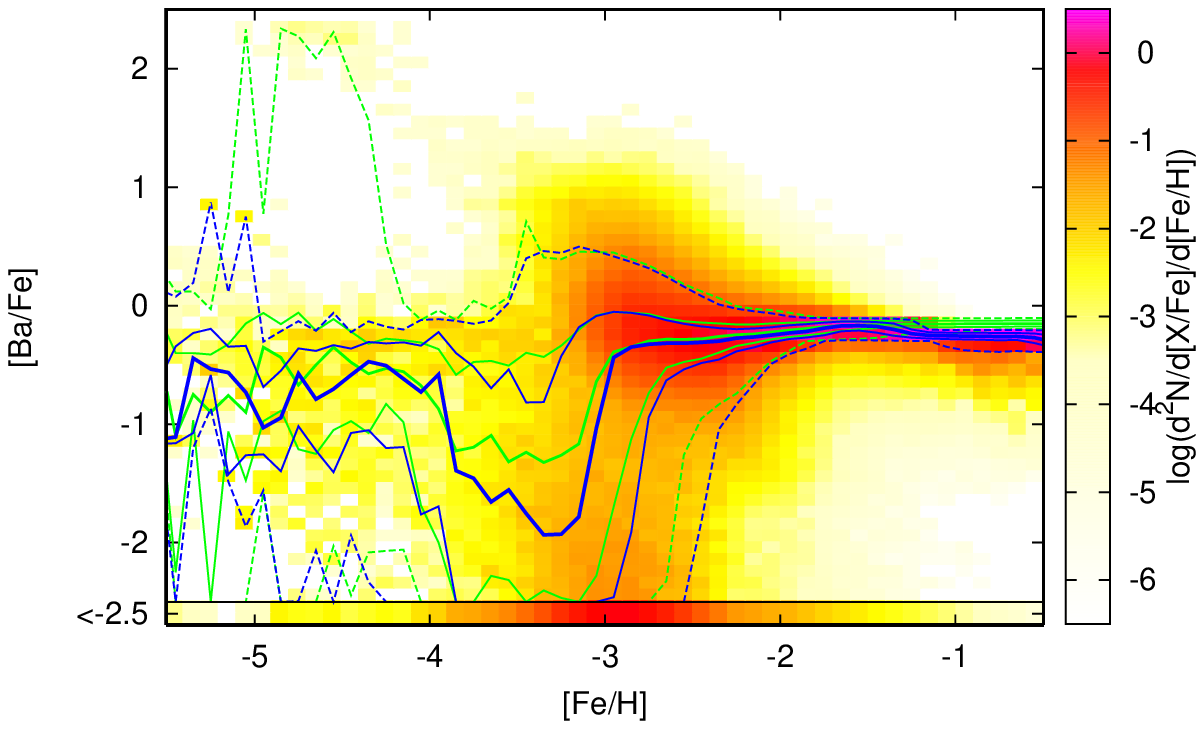}
\caption{
The predicted abundance distributions by our hierarchical models with the low-mass IMF of \citet{Chabrier03}. 
{\it Top panel:} Result with the same mass range, \rps yield and star formation efficiency as Model C, 
i.e., $(M_{\rm lo}, M_{\rm hi})= (9\msun,10\msun)$, $P=15.3\%$ and $\epsilon_\star=1\times10^{10} {\rm yr}^{-1}$.  
{\it Bottom panel:} Result with the same frequency of \rps source, yield and SN rate as Model C, 
i.e., $(M_{\rm lo}, M_{\rm hi})= (9\msun, 9.62\msun)$, $P=10\%$ and $\epsilon_\star=4.8\times10^{10} {\rm yr}^{-1}$.   
The distribution of \rps element abundance indirectly depends on the IMF through SN rate and $P$. 
}
\label{chabrierIMF}
\end{figure}

The total number of EMP survivors predicted is strongly dependent on the IMF. 
The low-mass IMF predicts much larger EMP survivors than observed (see Paper I and Paper II for details). 
As shown in Figure \ref{chabrierIMF}, however, the abundance distribution of $\abra{Ba}{Fe}$ results rather similar to the high-mass IMF cases. 
The top panel shows the result of Model S with the same $M_{\rm lo}$, $M_{\rm hi}$ and $\epsilon_\star$ as Model C. 

We note that the event rate of SN under the low-mass IMF is smaller by $\sim 1/5$ than under the high-mass one for the same SFE. 
Main differences of Model S from the fiducial model are the smaller scatter of $\abra{Ba}{Fe}$ and the weaker decreasing trend of abundance for $\feoh \gtrsim -1$. 
They are due to the lower frequency of SNe under the low-mass IMF, and hence, to the slower progress of chemical evolution. 
The predicted abundance distribution is more similar to Model C01 with lower star-formation efficiency since the SN rates result similar. 
   The other difference in the trend at high metallicity of $\feoh \gtrsim -1$ is due to the contribution from type~Ia SNe, since 
the low-mass IMF gives larger frequency of type~Ia SNe relative to type~II SNe than the high-mass IMF. 

Resultant total star formation rate for Model S is also plotted in Figure~\ref{SFR}. 
Due to the low SN rate, SFR is little affected by the galactic outflow, and becomes higher than Model C with high-mass IMF. 
The resultant SN rate in Model S is similar to Model C01, while most of stellar mass goes to low-mass stars, which results much more numerous EMP survivors. 

The bottom panel of Figure~\ref{chabrierIMF} shows the result of Model S10, for which we set $M_{\rm hi}$ and $\epsilon_\star$ to give the same $P$, $\Y{Ba}$, $\Y{Eu}$ and SN rate as the fiducial Model C. 
The predicted abundance distribution is almost same with Model C. 
The only significant difference is low $\abra{Ba}{Fe}$ at $\feoh \sim -3.5$. 
At this metallicity range, the accretion of ISM is dominant for the surface \rps element abundance. 
The accretion rate in the low-mass IMF models is smaller than in the fiducial model because of the binary composition among low-mass survivors. 
In the fiducial model, most EMP survivors are secondary companions in binary systems. 
For the secondary stars, the accretion rate is enhanced by the larger gravity of their primary companions. 
In Model S10, however, most EMP survivors are single stars or primary stars in binaries. 

In summary, the abundance distribution of EMP stars indirectly depends on the IMF through the SN rate and the number fraction of \rps source. 
In addition, the contribution of the ISM accretion is dependent on the IMF through the binaries. 

\red{
Although the total number of EMP stars is strongly dependent on the IMF of EMP stars, the IMF has little impact on the slope of MDF of EMP stars. 
The Pop. III IMF plays important role for the low-metallicity tail of the MDF.  
}

\section{Observational Constraints on R-process Element Enrichment}\label{obsS}

\subsection{Observational Data}\label{sampleS}

We use the SAGA database, which compiles the abundance data from the literature in which the abundances of EMP stars ($\feoh \leq -2.5$) are measured by spectroscopy \citep{Suda08, Suda11}.  
Therefore, the SAGA sample is strongly biased toward extremely metal-poor stars. 
For $\feoh \lesssim -2.8$, however, the sample can be regarded as unbiased because their \red{precise metallicity can be determined only by the high-resolution spectroscopy and we cannot choose more metal-poor stars at target selection for the high-resolution observations. 
SAGA has collected most published data with high resolution spectroscopy at $\feoh<-2.5$ and registere more than 400 EMP stars. 
}

We adopt the same sample criteria with Paper~II.
We leave out carbon-enhanced stars with $\abra{C}{Fe} > 0.7$ since their surface abundances are thought to be influenced by binary mass transfer. 
Especially, many carbon-enhanced stars show strong enhancement of {\it s}-process elements. 
We plot only high-resolution ($R\geq 20000$) sample in the figures. 
We use only giant samples since the detection limits of Ba and Eu for main-sequence turnoff stars are much higher than for giants. 

In the SAGA database, 216 giant EMP stars are registered leaving aside carbon-enhanced stars. 
The lowest abundance of the detected barium and europium are $\abra{Ba}{H} \sim -5.5$ and $\abra{Eu}{H} \sim -4$, respectively, for giant stars. 
It is thought that a significant number of stars are below the detection limit of europium. 
On the other hand, barium can be detected by high-resolution observations for most EMP giants as mentioned later. 

The SAGA database compiles data from many literatures. 
There can be systematic offsets of abundances due to differences in analysis methods between the literatures. 
For \rps elements, however, 
the intrinsic scatter is much larger than the systematic differences between the literatures. 
In Figure~\ref{fiducial}, \ref{param}, and following figures of $\feoh$ and $\abra{r}{Fe}$ plane, we plot observed samples by black symbols.  
Triangles$(\blacktriangle)$ denote the sample of the First Stars project \citep[][and 13 other papers of the series]{Hill02}.
  Squares ($\blacksquare$) denote the data of which the first author of a source paper is W.Aoki \citep[][and 15 other papers in the SAGA entry list]{Aoki02}. 
  Inverted triangle $(\blacktriangledown)$ shows the sample by \citet{Honda04, Honda07}. 
  Other samples are plotted with crosses ($+$). 

Recently, \citet{Andrievsky09} present the determinations of barium abundances taking into account the non local-thermal-equilibrium (NLTE) effects. 
The NLTE Ba abundances relative to Fe are slightly shifted toward the solar ratio. 
The slope of decreasing trend as metallicity decrease for $\feoh < -2.6$ is slightly reduced as the scatter. 
But the scatter remains quite large and the global trend is similar to the SAGA sample. 

\subsection{Comparison with Observations}
We make mock samples from the computation results in order to compare with observations. 
Since the SAGA sample is biased toward low metallicity and the MDF of SAGA sample at $\feoh > -2.8$ is almost flat.  We sample an equal number of stars from all the metallicity bins with $\Delta \feoh = 0.1$ at $\feoh>-2.8$. 
Then, we convolve the predicted abundances with a Gaussian observational error of $\sigma=0.2$dex. 
(For the color maps of abundance distributions, the Gaussian error is not convolved.)
In the following, we compare the\red{ predictions the observations about the median and scatter of \abra{$r$}{Fe}, the number fraction and metallicity of {\it r}-II stars ($\abra{Eu}{Fe}>1$) and the number of} stars with very low \rps abundances ($\abra{Ba}{H}<-5.5$) of mock samples. 
We give 90\% confidence intervals by bootstrap resampling and plot them as error bars in Figure~\ref{allresult}.

\subsection{Averaged Trend}\label{trendS}

We show the medians of $\abra{Ba}{Fe}$ at two metallicity ranges of $-2.8 < \feoh \leq -2.3$ and $-3.3 < \feoh \leq -2.8$ in Figure~\ref{allresult} and summarize the difference, $\Delta\abra{Ba}{Fe}_{\rm md}$, of them in Table~\ref{Tresult}. 
Median of predicted abundance of $\abra{Ba}{Fe}$ and $\abra{Eu}{Fe}$ are similar or slightly larger than observations in $-2.8 < \feoh \leq -2.3$ for all cases. 
This is a natural consequence of having set the \rps elements yields to give $\langle\abra{Ba}{Fe}\rangle=-0.1$ and $\langle\abra{Eu}{Fe}\rangle=0.6$. 

At higher metallicity, the predicted Ba abundance is lower than the observations since contributions from the {\it s}-process in intermediate massive stars are present in most stars, as pointed out by previous studies \citep[e.g.][]{Burris00}. 
\red{ In our fiducial model, the proto-galaxies become $\feoh \sim -2$ by 0.1 - 1 Gyr (see Figure~\ref{zFeH}), which is similar to the lifetime of intermediate-mass stars.  
}
The median of $\abra{Eu}{Fe}$ is also similar to the observed value at $\feoh > -2.3$ since the europium abundance is dominated by \rps throughout the metallicity range. 

On average, $\abra{Ba}{Fe}$ of observational sample shows a trend decreasing as the metallicity decreases in $-3.3 < \feoh < -2.3$, and our fiducial model reproduces the trend. 
For europium, the decreasing trend is obscured in the observed sample because of a higher detection limit. 
This decreasing trend strongly suggests that the dominant site of \rps elements is at the low-mass end of the SN mass range, as argued in some previous studies \citep[e.g.][]{Mathews92, Travaglio99, Qian08}.
Models with low-mass \rps sites predict a trend of decreasing $\abra{\it r}{Fe}$ as $\feoh$ decreases due to long delay time to the first event to provide \rps elements. 
The degree of the decreasing trend depends on $M_{\rm lo}, M_{\rm hi}$ and is consistent with the observations for $P\sim 3$--$10\%$.

\subsection{Scatter}\label{scatterS}

The distributions of both $\abra{Ba}{Fe}$ and $\abra{Eu}{Fe}$ of the observed EMP stars show large scatter. 
In our models, the predicted dispersion of $\abra{Eu}{Fe}$ is identical to $\abra{Ba}{Fe}$ since we assume a constant ratio, $\abra{Eu}{Ba}$, for the yield of all SNe. 
The scatter depends on the number fraction, $P$, of the \rps sources among SNe, as described in Section~\ref{rsiteS}. 
As seen in the middle panel of Fig.~\ref{allresult}, our hierarchical model in which the \rps elements originate from a limited mass range of progenitor stars with $P\sim 10$--$25\%$ can reproduce the observed scatter at the both metallicity ranges when we adopt the fiducial SFE. 

The abundance distributions also depend on the star-formation efficiency (SFE), $\epsilon_\star$, as mentioned in Section~\ref{resultSFES}.  
\red{ When we assume the high SFE of $10^{-9}$/yr, models with $P \sim 30 \%$ become compatible with the observed scatter at $\feoh \lesssim -2.5$. 
In this case, it is predicted that scatter at $\feoh \sim -2$ is also as large as at $\feoh \sim -2.5$. 
Observationally, 3 stars with $\abra{Eu}{Fe}>+1.5$ are identified but they cannot be reproduced in this case. 
}

\red{
When we assume the low SFE ($10^{-11}$/yr), the smaller mass range of $P = 1.5 \mhyph 2 \%$ becomes compatible with observations. 
}

In this paper, we assume that SN ejecta is mixed instantaneously and homogeneously in its host proto-galaxy, and when proto-galaxies merge, the element abundances are instantaneously averaged. 
The scatter of $\abra{\it r}{Fe}$ may be larger if the chemical inhomogeneity of proto-galaxies is taken into account. 
   For most proto-galaxies which birth EMP stars, however, the abundance homogeneity is thought to be a good approximation since their gas mass is comparable with typical masses of the gas swept up by SNe \citep{Machida05}.

\subsection{R-II Stars}\label{rIIS}
Among the 216 EMP giant stars in the SAGA sample, there are 12 {\it r}-II stars and three stars with $\abra{Eu}{Fe}>1.5$, which account for $5.6\%$ and $1.4\%$, respectively. 
It is known that iron abundance of {\it r}-II stars is concentrated around $\feoh = -2.8$ \citep{Barklem05}. 

Our fiducial model predicts that the number fraction, $f_{\rm rII}$, of {\it r}-II stars among EMP stars is $ 7.1\%$, and $1.3\%$ of EMP stars is $\abra{Eu}{Fe}>1.5$. 
The $90\%$ confidence interval of expected numbers of {\it r}-II stars is $10-22$ in 216 stars. 
These stars are formed in those low-mass proto-galaxies which experienced ECSNe at lower metallicity by chance. 

We also summarize the predicted $f_{\rm rII}$ and average metallicity of {\it r}-II stars in Table~\ref{Tresult}. 
For models with larger mass range, $P > 10\%$, it becomes smaller because of smaller $\Y{Ba}$, as seen for the bottom panel of Figure~\ref{allresult}.  
On the other hand, models with $P \lesssim 10\%$ predict $f_{\rm rII} \sim 7\%$ or smaller. 
For example, Model C01 with $P=1\%$ and a very large $\Y{Ba}$ gives $f_{\rm rII} = 4.2\%$. 
It is because the \rps site is very rare for Model C01 and takes place in more massive proto-galaxies than in the fiducial model; large $\Y{Ba}$ increases $f_{\rm rII}$ but larger $M_{\rm gas}$ decreases $f_{\rm rII}$.  
   The predicted fraction also depends on the SFE as shown in Figure~\ref{allresult}. 
The lower SFE models predict the lower $f_{\rm rII}$ because abundance of host proto-galaxies of EMP stars is averaged by merging at chemically earlier phase, as described in Section~\ref{resultSFES}.

The number fraction of {\it r}-II stars is also dependent on the explosion energy of a \rps source. 
Ejecta of explosion with large energy go away from the host halo as outflow. 
For small energy explosion, on the other hand, the yield can stay in the proto-galaxy even when the gas mass is very small and {\it r}-II stars are formed in such a galaxy with small mass. 
Observed large $f_{\rm rII}$ may indicate small explosion energy for \rps sources.  

Median, $\feoh_{\rm rII}$, of iron abundance of {\it r}-II stars is $\feoh = -2.91^{+0.14}_{-0.18}$ for the fiducial model. 
   The chemical abundance approaches to averaged value for higher metallicity while the number of stars decreases for lower metallicity. 
The predicted $\feoh_{\rm rII}$ is consistent with the observed metallicity of {\it r}-II stars.

\subsection{$R$-deficient EMP stars}\label{r0S}

Our fiducial model predicts that majority of stars with $\feoh < -3$ is \nor stars, which are formed without \rps elements. 
They are polluted with \rps elements by the ISM accretion but some of them remain with very low \rps abundances. 
The \rdemp survivors also contain some Pop.~III survivors and pre-enriched first stars which have undergone small ISM accretion. 

Among the 216 EMP giants in the SAGA sample, there are 184 stars with barium detected by spectroscopic observations, and only the upper limit is given to four stars. 
For the remaining 28 stars, there is no report about barium abundance. 
\red{ When we consider the stars with reported detection or upper limits of Ba abundance, the number fraction, $f_{r\rm df}$, of \rdemp stars is $4/188 \sim 2.1\%$. 
However, some of the stars without report of $\abra{Ba}{Fe}$ are thought to be very low barium abundances. 
For example, \citet{Barklem05} observed 125 EMP giant stars without carbon enhancement, and Ba is undetectable for 8 stars ($6.4\%$) among them (but they do not report upper limits). 
In Fig.~\ref{allresult}, we also plot total fraction of stars without the report of detected barium ($32/216 \sim 14.8\%$) as an upper bound of $f_{r\rm df}$. }

Our fiducial model (Model C) predicts $f_{r\rm df} = 9.5\% (13-28$ stars in the mock samples).  
\red{ This is compatible with the observations if about a half of stars without report of $\abra{Ba}{Fe}$ are \rdemp stars. }
The predicted value of $f_{r\rm df}$ is dependent on $P$ and $\epsilon_{\star}$, as shown in the bottom panel of Figure~\ref{allresult}. 
When $P$ is small, there are many proto-galaxies without \rps elements at the early stages of the Galaxy formation, and $f_{r\rm df}$ becomes large. 

The predicted value of $f_{r\rm df}$ is dependent on the accretion rate of ISM in our model, too. 
As shown in Figs.~\ref{accretion} and \ref{BaAcc}, in our fiducial model, most \nor stars become $\abra{Ba}{H}>-5.5$ by ISM accretion and only stars which accrete small quantity of barium remain as \rdemp survivors.  
The small percentage of \rdemp stars indicates that there is a significant contribution of surface pollution for most EMP survivors. 

\red{ \citet{Roederer13} argued that the stars for which Ba are not detected constitute only a small fraction. 
The ISM accretion is one possible scenario for the scarcity of \rdemp stars. }
There is another scenario that some secondary \rps site(s) with small yield of \rps element exists. 

\subsection{Mass-Dependent Yields}\label{argastS}
\begin{figure}[t]
\plotone{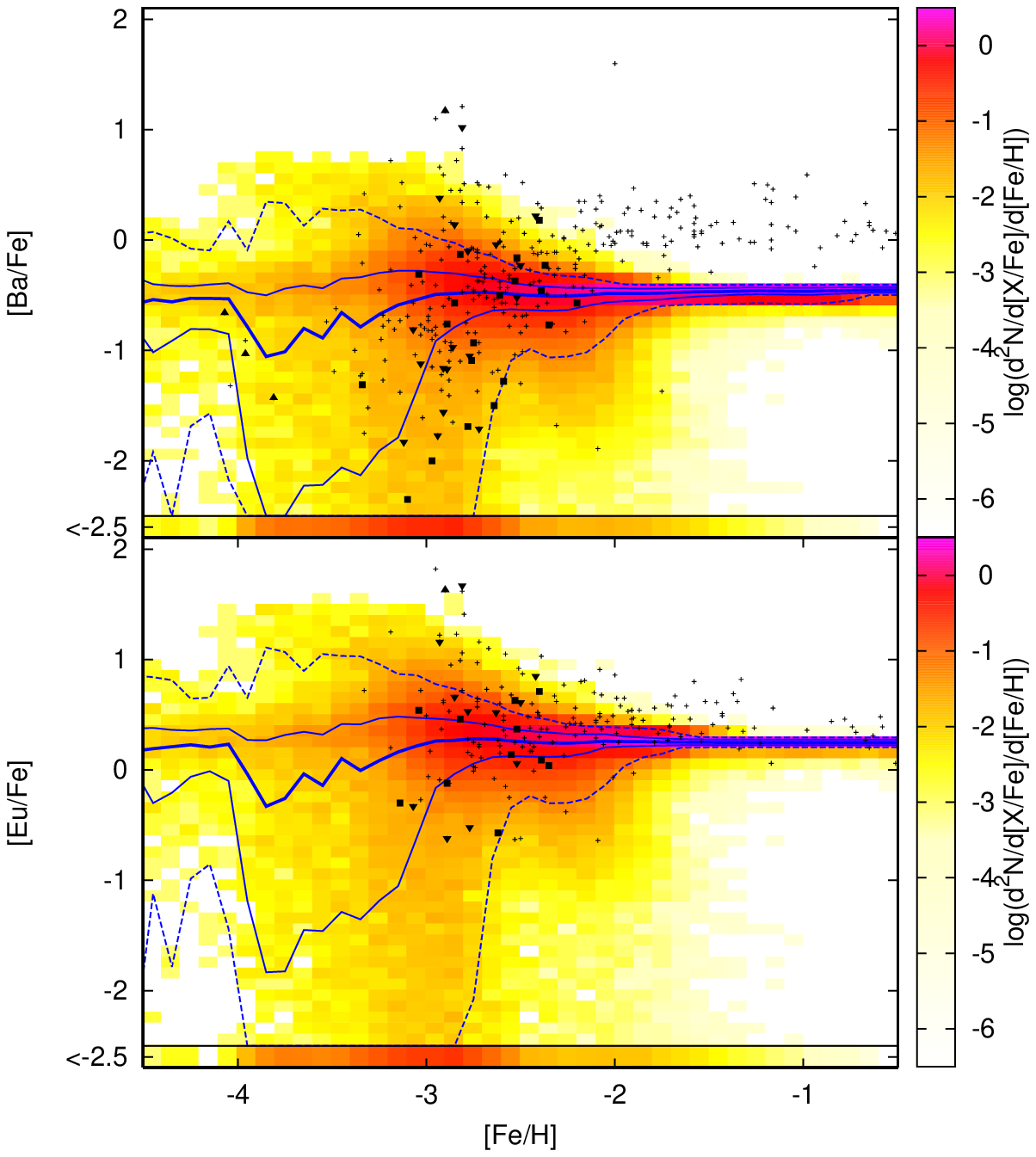}\label{argast}
\plotone{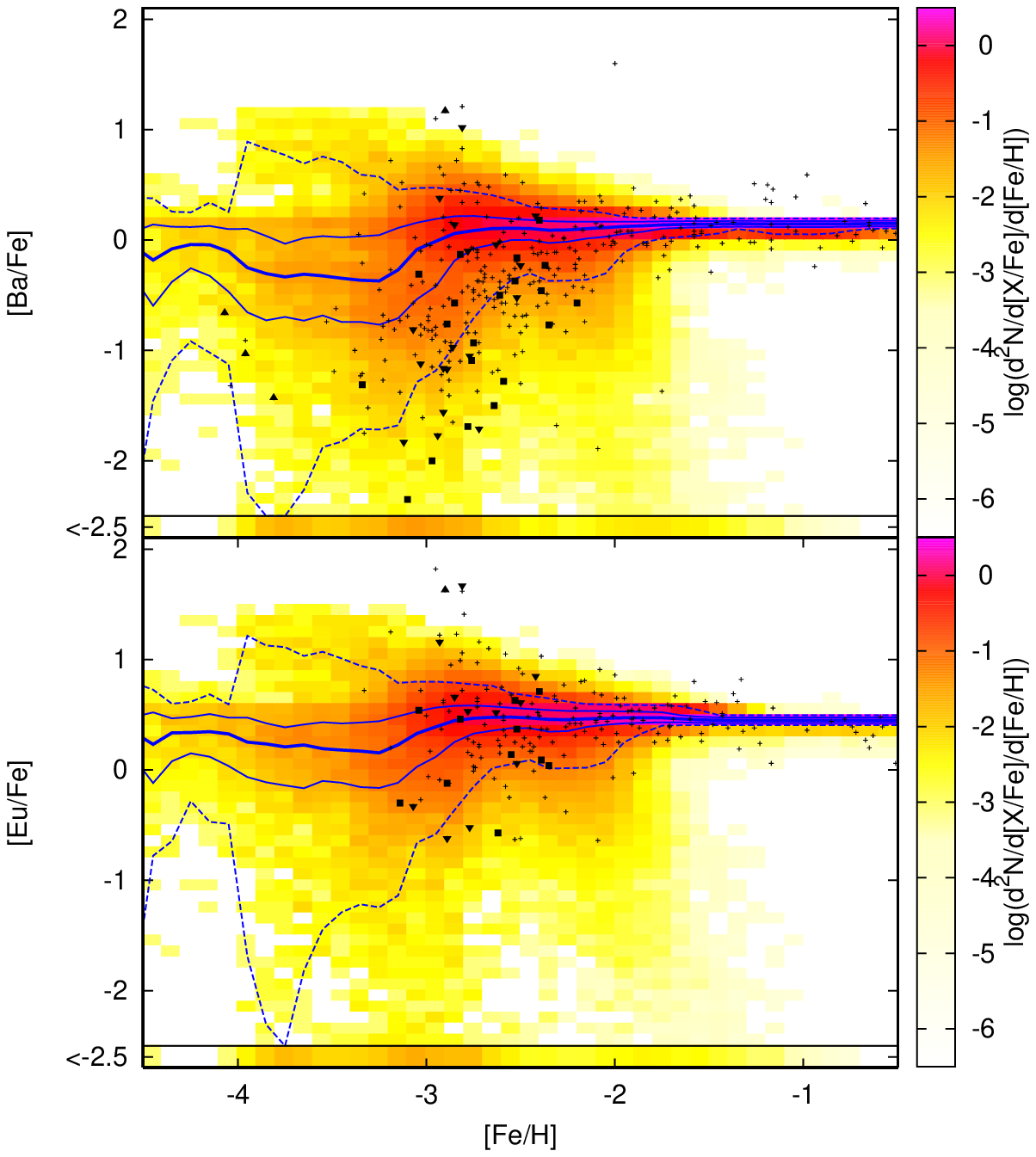}\label{cescutti}
\caption{The results with the \rps yields of \citet[][top panel]{Argast04} and \citet[][bottom panel]{Cescutti06}. }
\label{a+c}
\end{figure}

\citet{Argast04} and \citet{Cescutti06} proposed mass-dependent yield of Ba and Eu based on chemical evolution studies. 
For comparison, we compute our hierarchical chemical evolution model by adopting their \rps yields in Model SN2025 by \citet{Argast04} from SNe in the range between 20 and $25 \msun$ and in Model 1 of \citet{Cescutti06} from SNe in the range between 12 and $30 \msun$, both decreasing for larger mass.  
Figures~\ref{a+c} shows the results and the comparison with the observations.  
  
Both models predict shallower slopes of the decreasing trend of $\abra{Ba}{Fe}$ for lower metallicity in EMP stars since they assume as {\it r}-process sources SNe with higher mass progenitors than in our fiducial model. 
The scatters are also significantly smaller than observed and in our fiducial model (cf. Fig.~\ref{allresult}). 
In particular, the yield by \citet{Cescutti06} predicts a smaller scatter because of larger $P$. 
The predicted numbers are smaller for $r$-II stars than observed and the median of metallicity of $r$-II stars results lower.  
\red{But their yield can explain the observed very low fraction of \rdemp stars even when we consider only stars with upper limits for Ba. 
The discrepancy is partly due to the different assumptions on the gas infall and on the SFR from their study. 
It is somewhat alleviated when we adopt the smaller gas mass and the higher SFE at the early stage of chemical evolution as in \citet{Cescutti06}. 
}

\vspace{0.5cm}
In conclusion, $\sim 10\%$ (5--20\%) of CCSNe at the low-mass end of the SN mass range is a most plausible dominant site of r-process elements. 
In our hierarchical model, all the observational properties can be reproduced including $r$-II and \rdemp stars.

\section{Neutron Star Merger Scenario}\label{NSMS}

Neutron star (NS) merger has been proposed as a plausible dominant source of the \rps elements \citep{Freiburghaus99, Rosswog99}. 
Theoretical studies on the nucleosynthesis argue that NS - NS merger or NS - black hole merger can reproduce the observed abundance patterns of \rps elements \citep[e.g.][]{Wanajo12}.
On the other hand, \citet{Argast04} argued that NS merger is rejected as a dominant \rps source from the viewpoint of chemical evolution. 

We test the NS merger scenario for \rps elements in our hierarchical chemical evolution model \red{by the following formulation}. 
Under our fiducial assumptions of the IMF and binary parameters, $11\%$ of stars in the SN mass-range have secondary companions that become neutron stars ($9\msun<m_{\rm sec}<25\msun$). 
\red{The primary and secondary stars of binaries assumed to evolve independently and explode as SNe, respectively. }
Among these massive star binaries, we set the number fraction, $p_{\rm NSM}$, of those which form NS binaries and coalesce to eject \rps elements as $10$ -- $100 \%$.  
Under these assumptions with the high mass IMF, the event rate, $R_{\rm NSM}$, of NS merger becomes 
\begin{align}
R_{\rm NSM} & = & \psi \int^{25}_{9} \frac{dm_1}{m_1} \xi (m_1) \int^{m_1}_{9} dm_2 f_{\rm bin} {n(m_2/m_1)} p_{\rm NSM} \notag \\
 & \simeq &3.4 \times 10^{3} \left( \frac{\psi}{\msun {\rm yr}^{-1}} \right) p_{\rm NSM} {\rm Myr^{-1}}, 
\end{align}
where $f_{\rm bin}$ is a binary fraction and $n(m_2/m_1)$ is the mass ratio distribution of binaries. 
   This event rate is about ten times larger than the current Galactic event rate of NS mergers, estimated at $R_{\rm NSM} = 40 \mathchar`- 660 {\rm Myr}^{-1}$ from the pulsar observations \citep{Kalogera04} if we assume $\psi = 10 \msun {\rm yr}^{-1}$ and $p_{\rm NSM} = 0.1$, which may be ascribed to the high mass IMF assumed.  
   As shown in Sectin~\ref{resultIMFS}, the formation rate of massive stars for the high-mass IMF is $\sim 5$ times larger than for the IMF of local universe, and the number of NS binary is also larger. 
   The number fraction, $P$, of NS merge events per CCSNe, as defined in eq.~(\ref{eq:number_fraction}), is given by $P = 0.125 p_{\rm NSM}$.   
   The yield of \rps element is set to give $\langle\abra{Eu}{Fe}\rangle = 0.6$ as the same with the case of the CCSN scenario. 
	The adopted yields are $\Y{Ba}=5.72\times 10^{-5}$--$5.72\times 10^{-6}$ for $p_{\rm NSM} = 0.1$ and 1.0, respectively, comparable with the predicted yields by the numerical studies \citep[e.g.][]{Wanajo12, Bauswein13}. 

\begin{figure*}
\includegraphics[width=\textwidth]{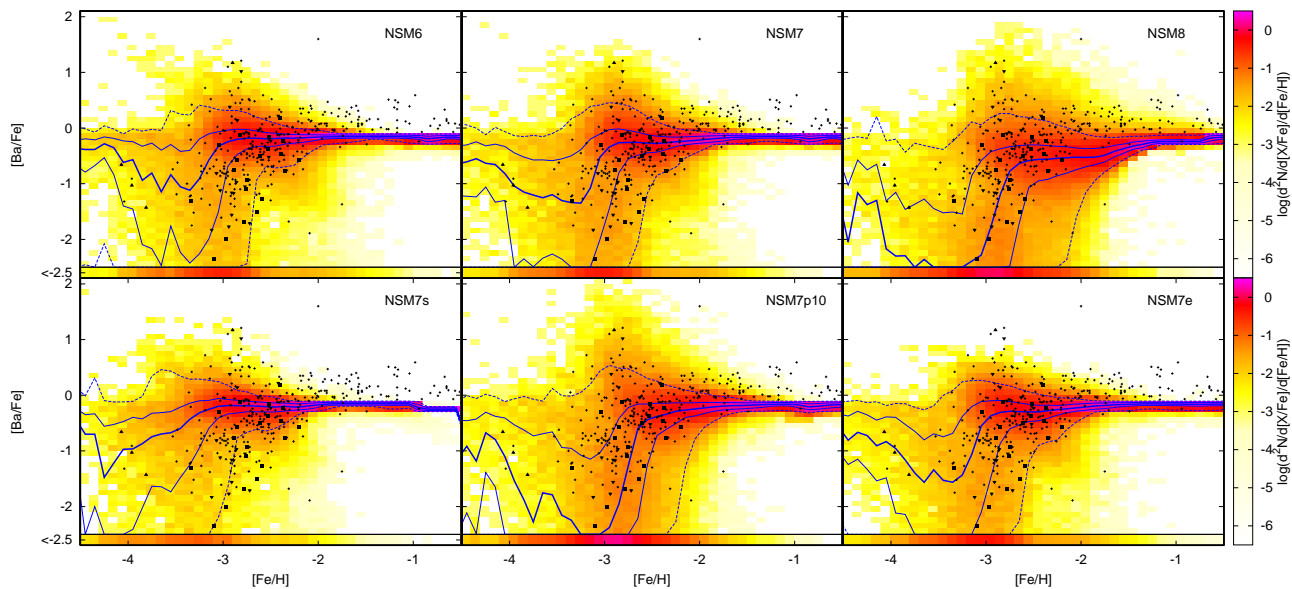}
\caption{Abundance distributions of \abra{Ba}{Fe} predicted from the NS merger scenario. 
{\it Top Left}:		Model NSM6 ($t_c=10^6$ yr and $p_{\rm NSM} = 100\%$.)
{\it Top Middle}:	Model NSM7 ($t_c=10^7$ yr and $p_{\rm NSM} = 100\%$.)
{\it Top Right}:	Model NSM8 ($t_c=10^8$ yr and $p_{\rm NSM} = 100\%$.) 
{\it Bottom Left}:	Model NSM7s ($t_c=10^7$ yr, $p_{\rm NSM} = 100\%$, $\epsilon_\star=10^{-11}{\rm yr}^{-1}$.)
{\it Bottom Middle}:Model NSM7p10 ($t_c=10^7$ yr and $p_{\rm NSM} = 10\%$.)
{\it Bottom Right}:	Model NSM7e ($t_c=10^7$ yr, $p_{\rm NSM} = 100\%$, $E_{\rm NSM}=10^{51}$erg.) .
Models assume $\epsilon_\star=10^{-10} {\rm yr}^{-1}$ and $E_{\rm NSM}=10^{50}$erg unless otherwise stated. 
}
\label{NSM}
\end{figure*}

\begin{figure}
\plotone{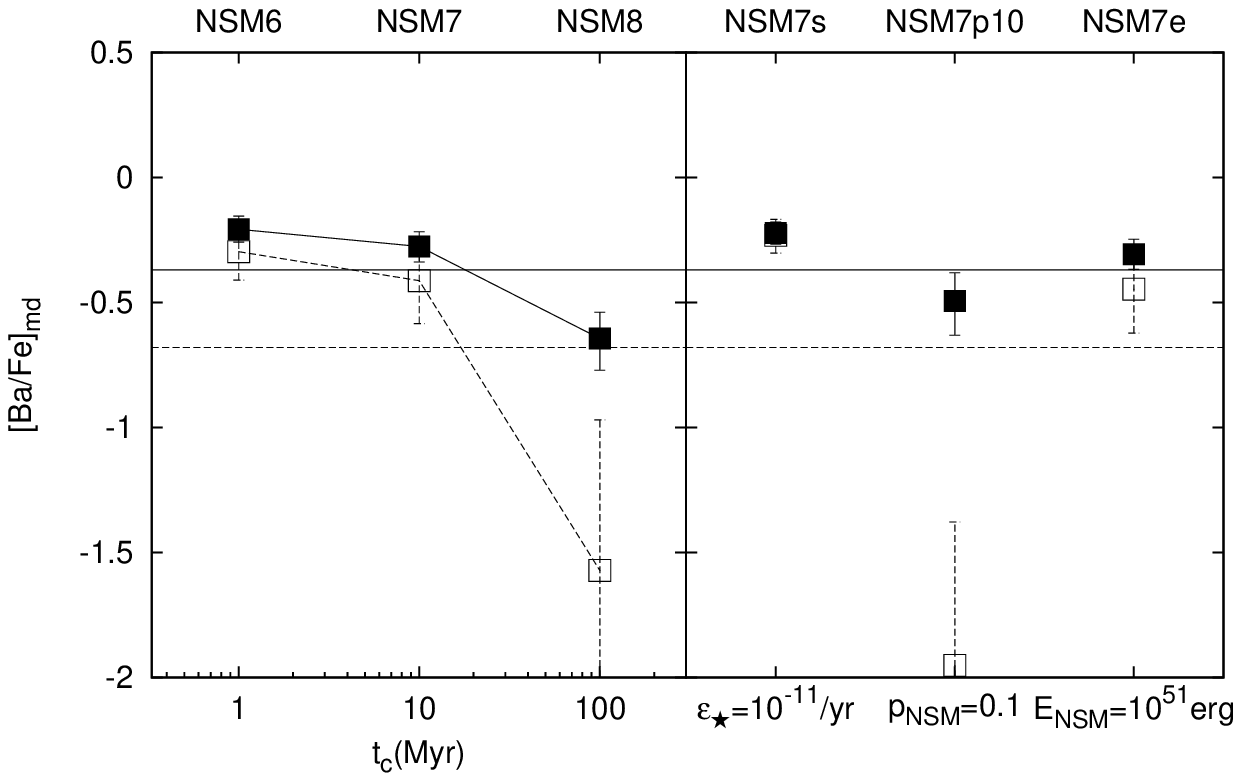}
\plotone{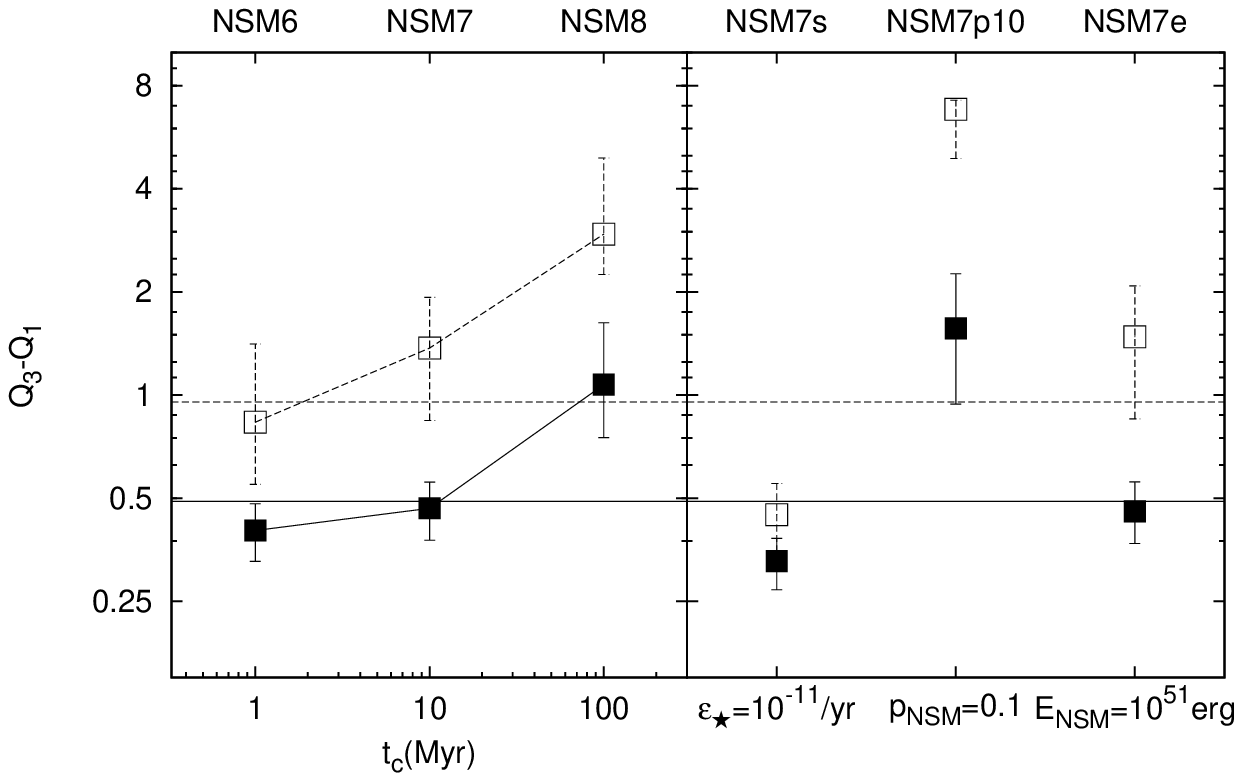}
\plotone{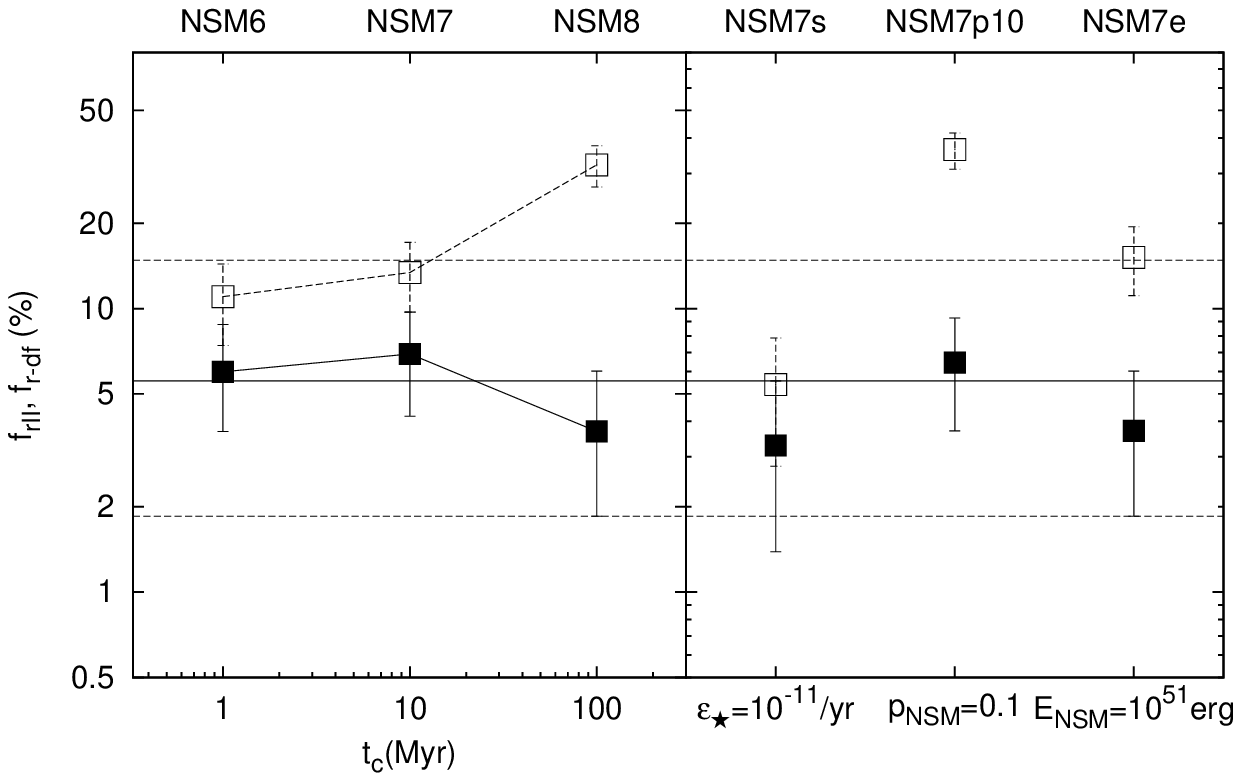}
\caption{ Same as Figure~\ref{allresult} but for neutron star merger scenario. 
}
\label{NSMall}
\end{figure}

The coalescence timescale for a NS binary is typically estimated to be $10^8$--$10^9$ yr \citep[e.g.][]{Fryer99}. 
\citet{Belczynski02} argue a much shorter timescale of around $10^6$ yr, from the consideration of common-envelope evolution, however. 
   In this study, we deal the timescale, $t_c$, \red{from the SN explosion of the secondary companion to the coalescence of the NS binary,} as a free parameter and set $t_c = 10^6 \mhyph 10^8$. 
\red{In order to estimate the typical timescale for the NS merger, we assume that all the NS binary coalesce in the same timescale of $t_c$. }
Explosion energy of the ejecta of the neutron star merger is another free parameter. 
   We set $E = 10^{50} \mhyph 10^{51}$ erg according to the numerical studies \citep[e.g.][]{Bauswein13, Hotokezaka13}. 

The parameters and results of models computed are given in Tables~\ref{Tmodel} and \ref{Tresult}, respectively.   
   Figure~\ref{NSM} shows and compares the predicted abundance with the observations for the models with the different assumptions on the coalescent timescale, event rate, and explosion energy.  
   Figure~\ref{NSMall} shows the parameter dependences of the results as in Figure~\ref{allresult}.  
   
Model NSN8 with $t_c=10^8$yr predicts that most EMP stars have very low \rps element abundances even after the surface pollution at variance with the observations. 
   Because of the long time delay, the metallicities of most proto-galaxies become $\feoh>-3$ before the coalescence of NS binaries and most EMP stars are formed without \rps elements. 
   Pollution rate by the ISM accretion is also too small because \rps abundance rises only after the host proto-galaxies merge for the majority of EMP stars. 
   This model predicts much larger fractions of \rdemp stars at variance with the observations, as seen from Fig.~\ref{NSMall}. 
   For Model NSN6 with $t_c=10^6$yr, on the contrary, NS mergers occur at lower metallicity.  
   There is a significant difference in the decreasing trend for the lowest metallicity range because of the contribution of NS mergers of massive progenitors in shorter timescales;  
   the median enrichment in $ -3.3 <\feoh \le -2.8$ is $\abra{Ba}{Fe}_{\rm md} = -0.29$, much larger than $-0.46$ of Model C, and also than observed.  

A model with $(t_c, p_{\rm NSM})=(10^7 {\rm yr}, 100\%)$ predicts the trend and scatter consistent with observations because of the similar delay time and event rate of \rps source to Model C.  
   For the smaller event rate of $p_{\rm NSM} = 0.1$, the models result in the smaller mean enrichment and larger scatters of $\abra{\it r}{Fe}$ abundances than observed for EMP stars, because the number fraction $P$ of \rps sources is very small as compared iron-producing CC-SNe and the \rps yield of a single event has to be very large.   
For the larger event rate of $p_{\rm NSM}=100\%$, on the other hand, the number fraction of \rps source results in $P = 12.5\%$, and the models behaves similarly to the fiducial Model C.  

Dependence on star formation efficiency is essentially the same as described in Section~\ref{resultSFES}. 
   The smaller SFE brings the smaller scatter of $\abra{Ba}{Fe}$ and smaller decreasing trend around $\feoh \sim -3$, as seen from Model NSM7s. 
\red{ If we adopt lower SFE, slightly smaller event rate with $p_{\rm NSM}\sim 0.3$ is compatible with observations, although small coalescence timescale ($\sim 10^7$ yr) is required in this case, too. } 

Model NSM7 assuming $E_{\rm NSM}=10^{50}{\rm erg}$ predicts the number fraction of $r$-II stars at $f_{r\rm II} = 6.9\%$, consistent with the observation.  
On the other hand, Model NSM7e with $E_{\rm NSM}=10^{51}{\rm erg}$ predicts smaller number fraction of $r$-II stars. 
  This may be due to gas outflow by a NS merger, since 
  $r$-II stars are formed in mini-halos of small gas masses. 
  In the case of $E_{\rm NSM}=10^{51}{\rm erg}$, \rps elements and gas in the host proto-galaxies are blown away from mini-halos, while \rps elements remain in proto-galaxies for smaller explosion energy.  
  
As a conclusion, the NS merger scenario can be consistent with the observed abundance distributions of heavy \rps elements differently from \citet{Argast04}, but constraints are imposed on the coalescence timescale and the event rate of NS mergers in the early evolutionary stages of the Milky Way formation.   
   It demands very short coalescence timescales of $t_c \simeq 10^7$ yr for most of mergers. 
   \red{If we adopt longer coalescence timescale, the NS merger scenario predict larger abundance scatter and much more \rdemp stars than observations, similar to the results obtained by \citet{Argast04}. }
  In addition, the event rate of NS mergers has to be very large, $\sim 100$ times larger than estimated in the present-day Milky Way. 
   Although a part of large event rate may be explicable in terms of the IMF, the differences in $t_c$ and $p_{\rm NSM}$ are likely to originate from the metallicity dependences of the common envelope evolution under the EMP circumstances. 

\section{Summary and Conclusions}\label{concludeS}

In this paper, we have studied the chemical evolution of \rps element abundance in the early universe in the framework of hierarchical formation of the Galaxy assuming the CCSNe and NS mergers as possible \rps sites.   
We investigate Ba and Eu abundance as representative of heavier \rps elements since the contribution of {\it s}-process is negligible not only for Eu but also for Ba at $\feoh \lesssim -2.3$. 
Our hierarchical chemical evolution model is based on the standard $\Lambda$CDM cosmology.
We follow the chemical evolution of $\sim100,000$ proto-galaxies, which are building blocks of the Galactic halo. 
Each individual EMP star is registered in the computations with its mass, which is set randomly following the IMF. 
One of the novelties of our model is to include the surface pollution of EMP survivors through the gas accretion. 
Another is to consider the inhomogeneous pre-enrichment of intergalactic medium (IGM) by the SN driven wind from proto-galaxies. 
Our model can reproduce the distributions of \rps element abundances observed for metal-poor stars as well as the metallicity distribution function. 
We have made quantitative comparisons between the model results and the observations for the median and scatter of \rps element abundances, the number and metallicity of $r$-II stars, and the percentage of \rdemp stars. 
   Based on the comparisons, we have derived the constraints both on \rps sites of CCSNe and NS mergers, necessary to reproduce the observed characteristics of EMP stars. 
Main conclusions are as follows. 

\begin{enumerate}
\item
The number fraction of the \rps sources should be $\sim 10\%$ of core-collapse supernovae (CCSNe) producing iron in order to explain the observed distribution of \rps element abundances of EMP survivors. 
The delay time in the enrichment of \rps elements after the iron enrichment should be a few tens of million years. 
The scenario assuming electron-capture supernovae (ECSNe) in the lowest mass range of type II SNe as the \rps sites satisfies these constraints. 
Our hierarchical model adopting the ECSN scenario well reproduces the observed large scatter among EMP stars, the decreasing trend of median enrichment for low metallicity, and the fractions of $r$-II and \rdemp stars. 
 The NS merger scenario can also reproduce the observed distributions in case that most NS binaries coalesce in short timescale of $\sim 10^7$ yr and the events rate is much larger by $\sim 100$ times than the currently observed Galactic rate. 

\item
Our hierarchical model predicts the chemical evolution tracks as follows: 
At first, the metallicity increases from $Z = 0$ to $-4\lesssim \feoh \lesssim-2.5$ with $\abra{\it r}{Fe}=-\infty$ by a PISN and/or an iron-core collapse SN. 
The first ECSN in the proto-galaxies cause jumps of barium abundance to $-3.5 \lesssim \abra{Ba}{H}\lesssim -2$. 
Diversity in the mass of proto-galaxies, the merging histories, and the mass of SN progenitors yield large dispersion of $\abra{\it r}{Fe}$. 
As the chemical evolution progresses and proto-galaxies merge, the scatter of the abundance is reduced. 
At $\feoh\gtrsim-1.5$, the scatter is smaller than the measurement errors. 

\item
Majority of stars at $\feoh<-3$ are formed without \rps elements. 
Their surfaces are then polluted with \rps elements through the accretion of interstellar medium (ISM). 
For stars with $\abra{Ba}{H}<-3.5$, the surface pollution plays a dominant role as long as their surface abundance is concerned. 

\item
Contribution from the pre-enrichment of proto-galaxies is marginal for abundance distribution of \rps elements. 
For the $10\%$ of EMP survivors, pre-enrichment is the dominant source of \rps elements in their interior and their averaged abundances are $\abra{Eu}{H}_{\rm i} \sim-3.6$. 
However, their intrinsic abundance is obscured by the surface pollution. 

\end{enumerate}

In the forthcoming paper, we will discuss the relative abundance of light and heavy \rps element, such as $\abra{Sr}{Ba}$.  
We will evaluate the contribution of light element primary process and importance of the surface pollution on the distribution of $\abra{Sr}{Ba}$. 

\begin{acknowledgements}
This work was supported by the JSPS KAKENHI Grant Number 25800115.
\end{acknowledgements}

\end{document}